\documentclass[pra,floatfix,amsmath,superscriptaddress,twocolumn]{revtex4}
\usepackage{amssymb}
\usepackage{graphicx}
\usepackage{graphics}
\usepackage{amsmath}
\usepackage{amsthm}
\usepackage{color}
\usepackage{mathtools}

\newcommand{\bea}{\begin{eqnarray}}
\newcommand{\eea}{\end{eqnarray}}

\def\bi{\begin{itemize}}
\def\ei{\end{itemize}}
\def\bc{\begin{center}}
\def\ec{\end{center}}

\def\C{\hbox{$\mit I$\kern-.7em$\mit C$}}
\def\R{\hbox{$\mit I$\kern-.6em$\mit R$}}
\def\N{\hbox{$\mit I$\kern-.6em$\mit N$}}

\def\ket#1{|#1\rangle}
\newcommand{\one}{\mbox{$1 \hspace{-1.0mm}  {\bf l}$}}

\def\ket#1{\left| #1\right>}
\def\bra#1{\left< #1\right|}

\newtheorem{theorem}{Theorem}
\newtheorem{corollary}[theorem]{Corollary}
\newtheorem{lemma}[theorem]{Lemma}

\newtheorem{observation}[theorem]{Observation}

\begin{document}

\author{David Sauerwein}
\affiliation{Institute for Theoretical Physics, University of
Innsbruck, Technikerstra{\ss}e 21a, A-6020 Innsbruck, Austria}
\author{Chiara Macchiavello}
\affiliation{Dip. Fisica and INFN Sez. Pavia, University of Pavia, via Bassi 6, I-27100 Pavia, Italy}
\author{Lorenzo Maccone}
\affiliation{Dip. Fisica and INFN Sez. Pavia, University of Pavia, via Bassi 6, I-27100 Pavia, Italy}
\author{Barbara Kraus}
\affiliation{Institute for Theoretical Physics, University of
Innsbruck, Technikerstra{\ss}e 21a, A-6020 Innsbruck, Austria}

\title{Pure state transformations via finitely many round LOCC protocols}

\title{Multipartite correlations in mutually unbiased bases}

\begin{abstract}
 We introduce new measures of multipartite quantum correlations based on classical correlations in mutually unbiased bases. 
 These classical correlations are measured in terms of 
the classical mutual information, which has a clear operational meaning. 
We present sufficient conditions under which these measures are maximized. States that have previously been shown to be particularily important in the 
context of LOCC transformations 
fulfill these conditions and therefore maximize the correlation measures. In addition, we show how the new measures 
can be used to detect high-dimensional tripartite entanglement by using only a few local measurements.
\end{abstract}

\maketitle

\section{Introduction}

The fact that quantum systems can exhibit correlations that have no classical analog is at the core of quantum theory.
In the last decades it has been realized that these quantum correlations are not only of fundamental interest; they can moreover 
be manipulated to achieve tasks that are impossible with classical devices. These applications include quantum communication, quantum computation and quantum simulation \cite{NielsenChuang}.
Hence, a lot of effort has been devoted to reach a better understanding of quantum correlations and to find new ways to detect, quantify and manipulate them. 
While the theory of bipartite quantum correlations is well developed, the multipartite case still holds many open problems \cite{horodecki, eltschka}.

The drastic difference between the bipartite and multipartite case is particularly clear in the case of quantum entanglement, which is fundamental to 
many applications in quantum information theory. 
For bipartite quantum systems there exists (up to local unitary operations) one single most entangled state, the maximally entangled state. 
Its name is justified by the fact that spatially separated parties can deterministically obtain any other bipartite state from the maximally entangled state by 
manipulating it via local operations assisted by classical communication (LOCC) \cite{nielsen}. The maximally entangled state is hence the optimal bipartite entanglement resource. It  
maximizes any measure of entanglement and serves as a standard with which the resourcefulness of other entangled bipartite quantum states can be compared. This situation changes drastically if one enters the
multipartite realm. Multipartite quantum systems can be entangled in many inequivalent ways \cite{duer} and generally there does not exist a unique maximally 
entangled state that can serve as a standard to assess the resourcefulness of other quantum states \cite{MES}. In the last years, several classes of
multipartite entangled quantum states with interesting properties have been identified and some of these have motivated novel multipartite applications, such as one-way quantum computing \cite{raussendorf}, 
secret sharing \cite{hillery} or applications in quantum metrology \cite{giovanetti}. But the correlations these states contain are nevertheless often only 
partially understood. A better understanding of their properties seems however crucial if one wants to find new truly multipartite applications of the 
quantum correlations they exhibit.

To this end, different approaches are being pursued. One approach is to qualify and quantify quantum correlations in terms of their transformation properties.
As entanglement is a resource under LOCC, it is natural to study LOCC transformations of multipartite quantum
states \cite{eltschka}. Moreover, more general transformations have been considered, such as separable transformations \cite{horodecki,chitambar} or 
$\epsilon$-nonentangling operations \cite{brandao}, despite the fact that they lack a clear physical meaning. Another approach is to find measures to quantify the 
quantum correlations contained in multipartite states. 
These include approaches based on, e.g. quantum entropic quantities, concurrences and polynomial invariants \cite{horodecki, eltschka}.

In recent work \cite{CompCorr,huang}, the concept of complementarity, another distinctive feature of quantum theory, has been used to quantify quantum correlations of bipartite quantum systems. In the present work, we generalize 
these results to the multipartite setting. Two observables are called complementary if certainty about the measurement 
outcome of one observable implies complete uncertainty about the outcome of the other observable. For complementary observables 
the absolute value of the overlap between any eigenstate of the first observable with any eigenstate of the second 
observable is constant \cite{Durt} and their eigenbases are called mutually unbiased.
Mutually unbiased bases (MUBs) have many applications ranging from quantum state tomography, quantum error correction, and quantum cryptography to 
the detection of entanglement \cite{Durt}.

In this work, we use the notion of complementary observables to study quantum correlations. The central idea is that only quantum correlated 
states can exhibit strong classical correlations in the measurement outcomes of local complementary observables. 
We use mutually unbiased bases to define a novel set of measures of quantum correlations, $\{\mathcal{C}_N\}$. 
$\mathcal{C}_N$ has a clear operational meaning as it measures multipartite quantum correlations in terms of the maximal classical mutual information that a 
single party can share with the other parties by measuring the state in $N$ mutually unbiased bases. This generalizes the bipartite measure of ``complementary correlations'' introduced in \cite{CompCorr} to the multipartite case. 
We investigate which states are maximally correlated with respect to $\mathcal{C}_N$ and present some necessary and some sufficient conditions for this 
to be the case, thereby providing new physical insight into the quantum 
correlations they contain. Moreover, we present several other applications of $\mathcal{C}_N$ for pure and mixed 
quantum states. One application is entanglement detection. While the effort required to certify entanglement generally increases rapidly with the system size \cite{guehne0}, we 
show how $\mathcal{C}_N$ can be used to detect even high-dimensional genuine tripartite entanglement employing only a few local measurement settings. The structure of the remainder of this paper is the following. 
In Sec. \ref{sec:PrelNot} we introduce our notation and recall important definitions and 
results on MUBs and entropic uncertainty relations. We moreover review some concepts from entanglement theory that are used in this work. After that we present the following results.
\begin{itemize}
\item[(i)] {\bf New correlation measures} (Sec. \ref{sec:newmeasure}):  We present a set of new correlation measures, $\{\mathcal{C}_N\}$, and discuss its basic properties.
\item[(ii)] {\bf States maximizing $\mathcal{C}_N$} (Sec. \ref{sec:MaxStates}): In Lemma \ref{lem:01} we first present a generalized version of a result from \cite{CompCorr} on bipartite states that maximize $\mathcal{C}_N$ 
(see also \cite{coles}).
In Lemma \ref{lem:1} we present necessary and in 
Theorem \ref{thm:2} sufficient conditions for a (multipartite) state to yield the maximal quantum correlations as measured by $\mathcal{C}_N$.
\item[(iii)] {\bf Examples of pure states maximizing $\mathcal{C}_N$} (Sec. \ref{sec:pure}): We use Theorem \ref{thm:2} to show that several previously studied few-body quantum states
maximize $\mathcal{C}_N$, 
discuss properties of $(n>3)$-partite pure states that maximize $\mathcal{C}_N$ and present examples thereof. Moreover, we investigate how $\mathcal{C}_N$ changes under local operations assisted by classical communication (LOCC).
\item[(iv)] {\bf Detection of mixed state entanglement} (Sec. \ref{sec:mixed}): We show how $\mathcal{C}_N$ can be used to detect genuine tripartite entanglement (Lemma \ref{lem:fsep} and Lemma \ref{lem:bisep}). 
In particular, we demonstrate that even for high-dimensional systems and using only two local measurement settings, $\mathcal{C}_N$ can be employed to detect 
genuine tripartite entanglement in the vicinity of the generalized GHZ state.
\item[(v)] {\bf Generalization to mutually unbiased measurements} (Sec. \ref{sec:genmeasurements}): We discuss how the definition of $\mathcal{C}_N$ can be generalized 
from measurements in MUBs to, e.g., include measurements in mutually unbiased measurements (MUMs) \cite{Gour1}.
\end{itemize}

\section{Preliminaries and Notation}
\label{sec:PrelNot}
In this subsection we introduce our notation and recall important results on MUBs and entropic uncertainty relations. Moreover, we review concepts from entanglement theory that we use in this work.

\subsection{Mutual information}
We denote by $\mathcal{D}(\mathcal{H})$ the set of density operators on the Hilbert space $\mathcal{H}$.
If a basis $\mathcal{B} = \{\ket{b(i)}\}_{i=0}^{d-1}$ of $\C^d$ is 
measured on $\rho \in \mathcal{D}(\C^d)$, this yields a specific outcome $i \in \{0,\ldots,d-1\}$, with probability 
$p(i|\mathcal{B};\rho)$. The uncertainty with respect to the outcome of the measurement is quantified by the
Shannon entropy of this probability distribution, which we denote by $H(\mathcal{B}|\rho) = - \sum_i p(i|\mathcal{B};\rho) \log(p(i|\mathcal{B};\rho))$. 
Here and in the following the logarithm is taken to base two. As we are only interested 
in the entropic properties of measurement outcomes, and not in the corresponding outcomes, we often identify an operator $X$ (no degeneracy) 
with its eigenbasis $\mathcal{B}$ 
and write e.g. $H(X|\rho) \equiv H(\mathcal{B}|\rho)$ and analogously for other quantities.
Moreover, for a state $\rho$ of a composite system of $n$ subsystems we measure the mutual dependence of the outcomes of measurements on different sets of subsystems via their mutual information.
More specifically, we consider sets of subsystems $A,B \subset \{1,\ldots,n\}$ that form a  partition of $\{1,\ldots,n\}$, i.e. $A \cap B = \emptyset$ and $A \cup B = \{1,\ldots,n\}$, and measurements in 
the basis $\mathcal{B}^{(l)}$ on subsystem $l$. The mutual information between the measurement outcomes in $A$ and $B$ is then 
\begin{align}
 &I(\{\mathcal{B}^{(l)}\}_{l \in A} : \{\mathcal{B}^{(l)}\}_{l \in B}|\rho)  \nonumber \\ 
 & = H(\{\mathcal{B}^{(l)}\}_{l \in A}|\rho) 
 - H(\{\mathcal{B}^{(l)}\}_{l \in A}| \{\mathcal{B}^{(l)}\}_{l \in B};\rho). \label{eq:MutInfo}
\end{align}
Here, we used the conditional entropy of the outcomes obtained at $A$ conditioned on the outcomes of $B$, 
\begin{align*}
 &H(\{\mathcal{B}^{(l)}\}_{l \in A}| \{\mathcal{B}^{(l)}\}_{l \in B}; \rho) \\
 &= H(\{\mathcal{B}^{(l)}\}_{l \in \{1,\ldots,n\}}| \rho) - H(\{\mathcal{B}^{(l)}\}_{l \in B}| \rho).
\end{align*}
Note here that $\mathcal{B}^{(l)}$ and $\mathcal{B}^{(l')}$ are allowed to be different for $l \neq l'$. 
If the global system is partitioned into one subsystem, $j$, and the rest, $\{1,\ldots,n\}/\{j\} \equiv \bar{j}$, we often use the notation 
$I(\mathcal{B}^{(j)} : \mathcal{B}^{(\bar{j})}|\rho)$ for the sake of readability and analogous abbreviations for other quantities. That is, $I(\mathcal{B}^{(j)} : \mathcal{B}^{(\bar{j})}|\rho)$ denotes the 
mutual information between subsystem $j$ and the rest when the bases $\{\mathcal{B}^{(l)}\}_l$ are measured. Let us denote the dimensions of the subsystems $A$ and $B$
with $D_S = \prod_{l \in S} d_l$ for $S \in \{A,B\}$.
Then the mutual information in Eq. (\ref{eq:MutInfo}) is upper bounded by $\min_A\{\log(D_A),\log(D_B)\}$. The bound is 
reached iff the outcomes of the measurements by one subset of parties $S \in \{A,B\}$ with $D_S = \min\{D_A,D_B\}$ are completely uncertain, but as soon as 
the measurement results of the other subsystems are known, one can predict them with certainty.

\subsection{Mutually unbiased bases}

Let us now review some definitions and results concerning mutually unbiased bases (see \cite{Durt} for a recent review). For a $d$-dimensional system a set of bases 
$\{\mathcal{B}_k\}_{k=1}^N$, where $\mathcal{B}_k = \{\ket{b_k(i)}\}_{i=0}^{d-1}$, is called a set of mutually unbiased bases (MUBs) if $\rvert\langle b_k(i)|b_{k'}(j)\rangle\rvert^2 = 1/d$ for all 
$k \neq k'$ and for all $i,j \in \{0,\ldots,d-1\}$. From the definition it is clear that any basis in a set of MUBs, $\{\mathcal{B}_k\}_{k=1}^N$, describes a 
property of the system's state that is complementary to the others. That is, knowing the outcome of a measurement in basis $\mathcal{B}_k$ that projects the quantum state onto $\ket{b_k(i)}$ implies that one has no 
prior knowledge about the outcome of a measurement in a complementary basis, 
i.e. $H(\mathcal{B}_{k'}|\ket{b_k(i)}\bra{b_k(i)}) = \log(d)$, for all $k' \neq k$. 

There exist at most $N = d+1$ MUBs in $\C^d$ \cite{Ivan0} and a set of two MUBs always exists \cite{Durt}. If the maximal number of $d+1$ MUBs 
is reached, the corresponding set is called complete.
A complete set of MUBs can be constructed if $d$ is prime or a power of a prime (see e.g. \cite{wootters,spengler0}). For the smallest dimension which is not of this kind, i.e. for $d = 6$, it is not known whether a complete set of MUBs exists.

In this work we often use that a particular complete set of MUBs is given by the eigenbases of some of the generalized 
Pauli operators if $d$ is prime. Their definition is as follows. For every $k = (k_1,k_2) \in \{0,\ldots,d-1\}^2$ a generalized Pauli operator is defined as 
\begin{align}
\label{eq:genpauli}
 S_{d,k} = X_d^{k_1} Z_d^{k_2},
\end{align}
where 
\begin{align*}
 &X_d = \sum_{k = 0}^{d-1} \ket{k+1 \ \text{mod} \ d}\bra{k},\\
 &Z_d = \sum_{k = 0}^{d-1} \omega_d^k \ket{k}\bra{k},
\end{align*}
and $\omega_d = \exp(2\pi i/d)$. For $d=2$ the operators in Eq. (\ref{eq:genpauli}) are proportional to the Pauli operators. The eigenbases of the set of 
generalized Pauli operators $\{S_{d,(0,1)}\} \cup \{S_{d,(1,m)}\}_{m=0}^{d-1}$ constitute a complete set of MUBs for $d$ prime \cite{bandyo}. The eigenvectors of $S_{d,k}$, $\{\ket{i_k}\}_{i=0}^{d-1}$, 
fulfill $S_{d,k}\ket{i_k} = \omega_d^i \ket{i_k}$. Notice also that the eigenbasis of $Z_d$ is the computational basis and the mutually unbiased Fourier basis
is the eigenbasis of $X_d$, i.e. $\ket{i_{(0,1)}} = \ket{i}$ and $\ket{i_{(1,0)}} = 1/\sqrt{d} \sum_{m=0}^{d-1} \omega_{d}^{mi} \ket{m}$ for $i \in \{0,\ldots,d-1\}$. 
Moreover, the eigenbases of $Z_d, X_d$ constitute a set of two MUBs for any $d$ \cite{Durt}.

\subsection{Entropic uncertainty relations}

The fact that each basis in a set of MUBs describes a 
property of the system's state that is complementary to the others is expressed in the well-known Maassen-Uffink-inequality \cite{Muffink}. For two arbitrary bases $\mathcal{B}'_1 = \left\{\ket{b'_1(i)}\right\}_{i=0}^{d-1}$ and 
$\mathcal{B}'_2 = \left\{\ket{b'_2(i)}\right\}_{i=0}^{d-1}$ it states that 
\begin{align}
H(\mathcal{B}'_1|\rho) + H(\mathcal{B}'_2|\rho) \geq -\log(c), \ \forall \rho \in \mathcal{D}(\mathcal{H}),
\end{align}
where $c = \max_{i,j} |\langle b'_1(i)|b'_2(j)\rangle|^2$. For a set of MUBs $\{\mathcal{B}_k\}_{k=1}^N$ this yields the entropic uncertainty relation 
\begin{align}
\label{eq:EUR0}
 H(\mathcal{B}_k|\rho) + H(\mathcal{B}_{k'}|\rho) \geq \log(d), \ \forall k \neq k'.
\end{align}
This relation shows that there is a tradeoff between the prior knowledge one can attain about the outcomes of measurements in basis $\mathcal{B}_k$ and $\mathcal{B}_{k'}$.

By simply applying inequality Eq. (\ref{eq:EUR0}) to always two different MUBs it is trivial to find the uncertainty relation
\begin{align}
\label{eq:EUR1}
 \sum_{k=1}^{N} H(\mathcal{B}_k|\rho) \geq \frac{N}{2} \log(d).
\end{align}
In \cite{Wehner} this relation was improved to 
\begin{align}
\label{eq:EUR2}
 \sum_{k=1}^{N} H(\mathcal{B}_k|\rho) \geq - N \log{\left( \frac{N + d - 1}{d N} \right)}.
\end{align}
As mentioned above, a complete set of $N = d+1$ MUBs is known to exist if $d$ is a power of two. The lower bound can then be further 
improved to \cite{S-Ruiz}
\begin{align}
\label{eq:EUR3}
 \sum_{k=1}^{d+1} H(\mathcal{B}_k|\rho) \geq  \frac{d}{2}\log{\frac{d}{2}} + \left(\frac{d}{2}+1\right) \log\left(\frac{d}{2}+1\right).
\end{align}

\subsection{Concepts from entanglement theory}
\label{sec:enttheory}

Let us now review some concepts from entanglement theory that we use in this work. 
Two states are called local-unitarily (LU)-equivalent if they can be converted into each other by applying local unitaries. Moreover, we denote the maximally entangled bipartite state by $\ket{\phi^+} \propto \sum_{i=0}^{d-1} \ket{i}\ket{i}$. The generalized GHZ state of $n$ subsystems with local dimension $d$ is denoted by  
$\ket{GHZ_{d,n}} \propto \sum_{i=0}^{d-1} \ket{i}^{\otimes n}$.

The first concept that we briefly review here are stochastic LOCC (SLOCC) classes. Two states $\ket{\psi}$ and $\ket{\phi}$ are in the same SLOCC class if $\ket{\psi}$ 
can be transformed into $\ket{\phi}$ and vice versa via LOCC with some finite probability of success. The states are then called SLOCC equivalent. The states that are SLOCC equivalent to $\ket{\psi}$ constitute its 
SLOCC class. It is known that there are only two different SLOCC classes of genuinely tripartite entangled three-qubit states \cite{duer}. The first class is 
represented by the GHZ state, $\ket{GHZ_{2,3}}$, and the second class is represented by the W state, $\ket{W} \propto \ket{100}+\ket{010}+\ket{001}$. There are, however, infinitely many SLOCC classes of genuinely 
multipartite entangled three-qutrit or four-qubit states \cite{verstr, briand}.

The second concept that we review here is the one of the maximally entangled set (MES) introduced in \cite{MES}. It is the set of all truely multipartite entangled states in a Hilbert space, $\mathcal{H}$, that cannot be obtained from any other 
LU-inequivalent state via LOCC. That is, the MES is the minimal set from which any other truely multipartite entangled state can be generated deterministically via LOCC. 
The MES is thus a generalization of the maximally entangled bipartite state to multipartite quantum systems. Note that,
while the MES of bipartite systems contains only $\ket{\phi^+}$, the MES of three-qubits contains 
already infinitely many states \cite{MES, Spee1, hebenstreit}.

Finally, we review the definitions of the source and accessible entanglement, which were introduced in \cite{EnMes}. For a quantum 
state $\rho \in \mathcal{D}(\mathcal{H})$ one can define the sets
\begin{align}
 &\mathcal{M}_s(\rho) = \{\sigma \in \mathcal{D}(\mathcal{H}) \ s.t. \ \sigma \xrightarrow{LOCC} \rho\}, \\
 &\mathcal{M}_a(\rho) = \{\sigma' \in \mathcal{D}(\mathcal{H}) \ s.t. \ \rho \xrightarrow{LOCC} \sigma'\},  
\end{align}
i.e. the set of states from (to) which $\rho$ can be obtained (transformed) via LOCC, respectively. For any measure $\mu$ on the state space one can measure their volumes 
$V_k(\rho) \equiv \mu(\mathcal{M}_k(\rho))$ for $k \in \{s,a\}$ and define the source and accessible entanglement as 
\begin{align}
 E_s(\rho) \equiv 1 - \frac{V_s(\rho)}{\sup_{\sigma} V_s(\sigma)} \ \text{and} \ E_a(\rho) \equiv \frac{V_a(\rho)}{\sup_{\sigma} V_a(\sigma)},
\end{align}
respectively. The former quantifies how easy it is to generate $\rho$ from other states via LOCC, while the latter quantifies the potentiality of $\rho$ to be 
converted to other states via LOCC. Due to their operational character, it is easy to show that they are indeed entanglement measures \cite{EnMes}. The source and 
accessible entanglement and generalizations thereof have been calculated and used to study and characterize few-body entanglement in \cite{EnMes,EnMes1}. Here, we compare these entanglement measures for some states with 
the new correlation functions we introduce in the next section.\\

\section{New correlation measures}
\label{sec:newmeasure}
In this subsection we introduce a set of new correlation measures for multipartite quantum systems based on measurements in mutually unbiased bases. We consider a $n$-partite quantum system 
with Hilbert space $\mathcal{H} = \bigotimes_{i=1}^n\C^{d_i}$ for which a set of $N \geq 2$ MUBs, $\{\mathcal{B}_k^{(l)}\}_{k=1}^{N}$, 
exists on each subsystem $l \in \{1,\ldots,n\}$. We then define the function 
\begin{align}
\label{eq:C1}
 C_N(\rho,\{\mathcal{B}_k^{(l)}\}_{k,l}) =  \frac{1}{n N} \sum_{k=1}^{N} \sum_{l=1}^n I(\mathcal{B}_{k}^{(l)} : \mathcal{B}_{k}^{(\bar{l})}|\rho),
\end{align}
and we introduce the correlation function
\begin{align}
\label{eq:C2}
 \mathcal{C}_N(\rho) = \max_{\{\mathcal{B}_k^{(l)}\}_{k,l}} C_N(\rho,\{\mathcal{B}_k^{(l)}\}_{k,l}).
\end{align}
Here we used the notation $\{\mathcal{B}_k^{(l)}\}_{k,l} \equiv \{\mathcal{B}_k^{(l)}\}_{k \in \{1,\ldots,N\}, l \in \{1,\ldots,n\}}$. In Eq. (\ref{eq:C2}) we optimize over all such sets of MUBs. Let us mention here again that $\mathcal{B}_k^{(l)}$ and $\mathcal{B}_k^{(l')}$ are allowed to differ for $l \neq l'$. That is, the bases on the different subsystems can be different. Moreover, all bases on any subsystem are mutually unbiased.

Let us first comment on some general properties of $\mathcal{C}_N$. 
Notice first that $\mathcal{C}_2$ is defined for any multipartite system as there always exist two MUBs on each subsystem. If a set of $L>2$ MUBs exists on each subsystem we obtain a whole class of correlation measures, 
$\{\mathcal{C}_N\}_{N = 2}^L$. As the maximal number of MUBs on subsystem $l$ is upper bounded by $d_l +1$ the number of functions in this class is $ L \leq \min_l{d_l} + 1$. 
In order to make the functions in $\{\mathcal{C}_N\}$
comparable to each other they are normalized such that they can reach at most 
$1/n \sum_{i=1}^n \log(\tilde{d}_i)$, where $\tilde{d}_i = \min\{d_i,\prod_{l \in \bar{i}} d_l\}$, which is independent of $N$ and saturated iff all of the mutual informations in Eq. (\ref{eq:C1}) are maximized. 
In this work we consider only systems for which $\tilde{d}_i = d_i$ for all $i \in \{1,\ldots,n\}$ such that the maximal value of the functions $\{\mathcal{C}_N\}$ is $1/n \sum_{i=1}^n \log(d_i)$. Moreover, note that for $N > N'$ every state maximizing 
$\mathcal{C}_N$ also maximizes $\mathcal{C}_{N'}$.

For bipartite systems $\mathcal{C}_2$ coincides up to normalization with the entropic measure of ``complementary correlations'' introduced in \cite{CompCorr}, where it has been shown that only 
entangled bipartite states can exhibit correlations in mutually unbiased bases that are strong enough to exceed a certain value of $\mathcal{C}_2$. In Sec. \ref{sec:MaxStates} we review these results and show how they can be extended 
to the multipartite case. 
In the subsequent sections we investigate the set of correlation measures $\{\mathcal{C}_N\}$ in detail and use them to study correlations of multipartite pure and mixed states.
We identify necessary properties of states to maximize $\mathcal{C}_N$ and present also conditions that are 
sufficient for maximization. Moreover, we use $\{\mathcal{C}_N\}$ to study the entanglement properties of multipartite quantum states. We show, for example, how it can be used to detect even high-dimensional genuine
tripartite entanglement using only few measurement settings.

\section{States maximizing $\mathcal{C}_N$}
\label{sec:MaxStates}
In this section we determine some properties a state $\rho \in \mathcal{D}(\bigotimes_{i=1}^n \C^{d_i})$ necessarily has to fulfill in order to reach the upper 
bound $\mathcal{C}_N(\rho) = 1/n \sum_{i=1}^n \log(d_i)$. Moreover, we 
present conditions that are sufficient to reach this bound. We first consider bipartite states and then proceed with a study of the multipartite case.

\subsection{Bipartite states}
\label{sec:MaxStatesBip}

In this subsection we extend the result that maximally entangled states are the only states in $\mathcal{D}(\C^d \otimes \C^d)$ that maximize $\mathcal{C}_2$ (see \cite{CompCorr}) to a more general 
situation (see also \cite{coles}). In order to do so, we use Holevo's Theorem \cite{Holevo}, which we review here for the sake of readability.\\

Let us consider the following bipartite scenario with parties Alice ($A$) and Bob ($B$) (see also \cite{NielsenChuang}). Suppose Alice
encodes the random variable $X \in \{1,\ldots,m_A\}$ using the ensemble $\{p_X(i), \rho_X(i)\}_{i=1}^{m_A}$ of quantum states and Bob performs a measurement
corresponding to the positive-operator valued measure (POVM) 
elements $\{Q_Y(i)\}_{i=1}^{m_B}$ on this ensemble to obtain the random variable $Y \in \{1, \ldots, m_B\}$. Then Holevo's theorem states that for any such measurement Bob may perform,
\begin{align}
\label{eq:Holevo}
 I(X:Y) \leq S(\rho) - \sum_i p_X(i) S(\rho_X(i)),
\end{align}
where $S(\rho) \equiv -\text{tr}(\rho \log(\rho))$ is the von Neumann entropy and $\rho = \sum_i p_X(i) \rho_X(i)$.\\

Here, we consider the case where Alice and Bob initially share a quantum state $\rho_{AB} \in \mathcal{D}(\C^{d'} \otimes \C^d)$, with $d' \geq d$. Alice can measure her system in order to create an ensemble on Bob's system, which encodes a random variable as described above. In order to encode the random variable $X$ she performs the measurement corresponding to the POVM elements $\{P_X(j)\}$ and thereby creates the ensemble $\{p_X(i),\rho^{(B)}_{X}(i)\}$ on Bob's system \footnote{Note that the number of indices $i$ and $j$ do not have to coincide, as several outcomes of Alice could lead to the same state of Bob's system.}. In order to encode $Z$ she performs $\{P_Z(j')\}$ and thereby prepares $\{p_Z(i'),\rho^{(B)}_{Z}(i')\}$ on Bob's system. Moreover, Alice prepares the ensembles in such a way that they fulfill the following additional condition.
For the measurements $\{Q_{Y_X}(i)\}$ and $\{Q_{Y_Z}(i')\}$ on Bob's system that allow him to maximize $I(X:Y_X)$ and $I(Z:Y_Z)$, respectively, it holds that 
\begin{align}
\label{eq:noinfo}
 \text{tr}(Q_{Y_X}(i)\rho^{(B)}_{Z}(j')) = \text{tr}(Q_{Y_Z}(i')\rho^{(B)}_{X}(j)) = 1/d
\end{align}
for all $i,j,i',j'$. That is, the measurement that allows Bob to extract maximal information on $X$ ($Z$), i.e. to maximize $I(X:Y_X)$ ($I(Z:Y_Z)$), cannot be used to extract any information about the other 
random variable, $Z$ ($X$), respectively, as then all measurement outcomes are equally likely. However, if Alice informs Bob about which one of the two random variables she has used for the encoding, e.g. by sending one classical bit, Bob can extract the maximal information about the corresponding random 
variable. Note, however, that $I(X:Y_X) + I(Z:Y_Z) \leq 2 \log(d)$. Using Holevo's Theorem we prove the following lemma (see also \cite{CompCorr, coles}).

\begin{lemma}
\label{lem:01}
 For a bipartite state $\rho_{AB} \in \mathcal{D}(\C^{d'}\otimes \C^{d})$, with $d' \geq d$, there exist measurements as described before for which 
 $I(X:Y_X) + I(Z:Y_Z) = 2 \log(d)$ iff $\rho_{AB}$ admits the decomposition  
\begin{align}
 \rho_{AB} = \sum_k \tilde{q}_k (V_k \otimes \one)\ket{\phi^+}\bra{\phi^+}(V_k^\dagger \otimes \one), \label{eq:LOtomaxent}
\end{align}
where $\{\tilde{q}_k\}$ are probabilities, $V_k$ are isometries mapping $\C^d$ to $\C^{d'}$ and where $V_k$ and $V_{k'}$ have orthogonal images, i.e. $V_k^{\dagger}V_{k'} = 0$, for $k \neq k'$.
\end{lemma}

It is easy to see that the state $\rho_{AB}$ can be reversibly transformed into the maximally entangled state $\ket{\phi^+}$ via local operations of Alice. $\rho_{AB}$ hence contains the same entanglement as $\ket{\phi^+}$.

For the sake of readability, we give here only an outline of the proof, which is presented in Appendix A in full detail. 
It follows directly from Holevo's Theorem that $I(X:Y_X) + I(Z:Y_Z) = 2 \log(d)$ can be achieved only if
\begin{align}
\label{eq:compmixedmain}
 S\left(\rho^{(B)}\right) = \log(d)
\end{align}
and
\begin{align}
\label{eq:puremain}
 S\left(\rho^{(B)}_{R}(i)\right) = 0, \ \forall i \ \text{and for} \ R \in \{X,Z\}
\end{align}
where $\rho^{(B)} = \text{tr}_A(\rho_{AB}) = \sum_{i} p_{R}(i) \rho^{(B)}_{R}(i)$, for $R \in \{X,Z\}$.
Eqs. (\ref{eq:compmixedmain} - \ref{eq:puremain}) state that 
$I(X:Y_X) + I(Z:Y_Z) = 2 \log(d_B)$ only if $\rho^{(B)}$ is completely mixed and Alice can prepare pure states on Bob's system using any one of the measurements. 
 Moreover, using Eq. (\ref{eq:noinfo}) it is easy to see that the pure state ensembles created on Bob's system have to 
correspond to MUBs. In Appendix A we show that these conditions can only be fulfilled if $\rho_{AB}$ can be expressed as described in Eq. (\ref{eq:LOtomaxent}).
 It is straightforward to see that the operations needed to transform $\rho_{AB}$ to the maximally entangled state can always be included in the measurements of Alice. 
Hence, it is enough to show the ``if''-part for the state $\ket{\phi^+}$, 
which has also been shown in \cite{CompCorr,coles}. The maximal correlation is reached if Alice and Bob 
measure either both in the computational basis, i.e. the eigenbasis of $Z_d$, or the mutually unbiased Fourier basis, i.e. the eigenbasis of $X_d$. A straightforward calculation shows that these measurements and the ensembles that Alice creates indeed fulfill Eq. (\ref{eq:noinfo}).

We can now look at the more special situation in which the two measurements by Alice are restricted to MUBs. The measurement setting described before Lemma \ref{lem:01} is then identical to the scenario considered in the definition of $\mathcal{C}_2$. For $d' = d$ it is easy to see that no mixed state of the form given in Eq. (\ref{eq:LOtomaxent}) exists. Any state that maximizes $\mathcal{C}_2$ thus has to be LU-equivalent to $\ket{\phi^+}$. 
We therefore obtain the following corollary of Lemma \ref{lem:01} (see also \cite{CompCorr,coles}).
\begin{corollary}
\label{cor:1}
 A bipartite state $\rho \in \mathcal{D}(\C^d \otimes \C^d)$ maximizes $\mathcal{C}_2$ iff it is LU-equivalent to the maximally entangled state, $\ket{\phi^+}$.
\end{corollary}

Since, as mentioned before, any bipartite state that maximizes $\mathcal{C}_N$ maximizes $\mathcal{C}_2$ as well, we have that any bipartite state maximizing $\mathcal{C}_N$ for $d' = d$ necessarily has to be pure and LU-equivalent to $\ket{\phi^+}$. However, it is easy to find examples of mixed states of systems with $d' > d$ that maximize $\mathcal{C}_2$ using Eq. (\ref{eq:LOtomaxent}). In the following 
section we derive some necessary and some sufficient conditions a multipartite state has to fulfill in order to maximize $\mathcal{C}_N$.

\subsection{Multipartite states}
\label{sec:MaxStatesMult}

Let us first derive some necessary conditions a state $\rho \in \mathcal{D}(\bigotimes_i^n \C^{d_i})$ has to fulfill in order to reach the upper bound
$\mathcal{C}_N(\rho) = 1/n \sum_{i=1}^n \log(d_i)$. Before that, recall that 
in order to determine $\mathcal{C}_N$ all systems are measured in $N$ different MUBs and for each of these measurement settings the mutual information between the measurement results obtained for any single system and
the rest are considered (see Eq. (\ref{eq:C1} - \ref{eq:C2})). Using Lemma  
\ref{lem:01} it is easy to show the following lemma.
\begin{lemma}
\label{lem:1}
 A multipartite state maximizes $\mathcal{C}_N$ only if it admits a decomposition as described in Eq. (\ref{eq:LOtomaxent}) in the bipartition of any single subsystem and the rest.
\end{lemma}
\proof{ Note that $\rho$ maximizes $\mathcal{C}_N$ only if it maximizes also $\mathcal{C}_2$. 
From the definition of $\mathcal{C}_2$ we have that $\rho$ maximizes this correlation function only if it maximizes the corresponding function $\mathcal{C}_2$ in each bipartite splitting of subsystem $l \in \{1,\ldots,n\}$ with the rest. It follows from the definition of $\mathcal{C}_2$ and Lemma \ref{lem:01} that this is possible only if $\rho$ admits a decomposition as described in Eq. (\ref{eq:LOtomaxent}) in the bipartition of any single subsystem and the rest. This proves the statement. \qed\\
}

We have stated in Corollary \ref{cor:1} that for bipartite systems with equal local dimensions, only pure states can maximize $C_N$. In contrast to that, we have the following observation for multipartite systems. 
\begin{observation}
 There exist mixed multipartite states that maximize $\mathcal{C}_N$, even if all subsystems have the same dimension.
\end{observation}

An example is the mixed three-qutrit state $\rho_{ABC} = \text{tr}_{R}(\ket{\Omega}_{RABC}\bra{\Omega})$, where 
\begin{align}
\ket{\Omega}_{RABC} = \frac{1}{3} \sum_{i,j = 0}^2 \ket{i}_R\ket{j}_A\ket{i+j \ \text{mod} \ 3}_B \ket{i+2j \ \text{mod} \ 3}_C \label{eq:AME}
\end{align}
is an absolutely maximally entangled state (AMES) presented in \cite{AMES}.  AMES are states of $N$-subsystems for which any $\lfloor N/2 \rfloor$-subsystem reduced state is completely mixed (see e.g. \cite{AMES}), where $\lfloor \cdot \rfloor$ denotes the floor of a number. One can show that $\mathcal{C}_N(\rho_{ABC}) = \log(3)$, for $N \in \{2,3,4\}$, which is the maximal value for all $N \in \{2,3,4\}$. These values are achieved if the individual subsystems are all measured in the basis of the same three-qutrit generalized Pauli operators.\\

The next theorem provides a condition that is sufficient to show that a state maximizes 
$\mathcal{C}_N$. Before we state this theorem, we introduce the following definitions. 
We denote by 
$GL(\C,d)$ the set of invertible operators on $\C^d$ and define the stabilizer of a pure multipartite state $\ket{\psi} \in \bigotimes_{i=1}^n \C^{d_i}$ as 
\begin{align}
 S_{\psi} = \{S \in \bigotimes_{i=1}^n GL(\C,d_i) \ s.t. \ S\ket{\psi} = \ket{\psi}\},
\end{align}
i.e. as the set of all local symmetries of $\ket{\psi}$. Using this notation, we state the following theorem (recall the definition of $S_{d,k}$ in Eq. (\ref{eq:genpauli})).

\begin{theorem}
\label{thm:2}
Let $\ket{\psi} \in (\C^d)^{\otimes n}$ be a pure state with the following properties.
\begin{itemize}
 \item $\ket{\psi}$ has completely mixed single-subsystem reduced states, i.e. $\text{tr}_{\bar{l}}(\ket{\psi}\bra{\psi}) \propto \one, \forall l \in \{1,\ldots,n\}$, and
 \item $\{\bigotimes_{l=1}^n S_{d,k}^{m_{k,l}}\}_{k \in \mathcal{K}} \subset S_{\psi}$, where $\{m_{k,l} \neq 0\}_{k,l} \subset \N$, and 
 $\mathcal{K} \subseteq \{0,\ldots,d-1\}^2, |\mathcal{K}| = N$, is such that the corresponding 
 set of generalized Pauli operators, $\{S_{d,k}\}_{k \in \mathcal{K}}$, is mutually unbiased.
\end{itemize}
 Then every $\ket{\phi} \simeq_{LU} \ket{\psi}$ maximizes $\mathcal{C}_N$.
\end{theorem}

\proof{It is clear that $\ket{\phi} \simeq_{LU} \ket{\psi}$ maximizes $\mathcal{C}_N$ iff $\ket{\psi}$ does. Hence, in order to proof the theorem it is sufficient to show that it holds for $\ket{\psi}$ itself.

We expand $\ket{\psi}$ on the first $n-1$ subsystems in the eigenbasis of $S_{d,k}$, $\{\ket{i_k}\}_{i=0}^{d-1}$, where $k \in \mathcal{K}$, as 
\begin{align}
\label{eq:deceig}
 \ket{\psi} = \sum_{\vec{i} \in \{0,\ldots,d-1\}^{n-1}} \ket{\vec{i}_{k}} \ket{\phi_{k}(\vec{i})}.
\end{align}
Here, we used the notation $\vec{i} = (i^{(1)},\ldots,i^{(n-1)})$ and $\ket{\vec{i}_{k}} \equiv \ket{{i^{(1)}}_{k}}\otimes \ldots \otimes \ket{{i^{(n-1)}}_{k}}$ is a state of the first $n-1$ subsystems.
As $\bigotimes_{l=1}^n S_{d,k}^{m_{k,l}} \in S_{\psi}$, we know that the equation $\bigotimes_{l=1}^n S_{d,k}^{m_{k,l}} \ket{\psi} = \ket{\psi}$ has to hold. Using Eq. (\ref{eq:deceig}) and that 
$S_{d,k} \ket{i_k} = \omega_d^i \ket{i_k}$ we obtain
\begin{align}
 \bigotimes_{l=1}^n S_{d,k}^{m_{k,l}}\ket{\psi} &= \sum_{\vec{i}} \ket{\vec{i}_{k}}  \omega_d^{\sum_{j=1}^{n-1} m_{k,j} i^{(j)}} S_{d,k}^{m_{k,n}}\ket{\phi_{k}(\vec{i})} \nonumber \\
 &= \sum_{\vec{i}} \ket{\vec{i}_{k}} \ket{\phi_{k}(\vec{i})} = \ket{\psi}. \label{eq:applysymm}
\end{align}
This equation is fulfilled iff
\begin{align}
\label{eq:EV33_2}
S_{d,k}^{m_{k,n}}\ket{\phi_{k}(\vec{i})} =  \omega_d^{-\sum_{j=1}^{n-1} m_{k,j} i^{(j)}} \ket{\phi_{k}(\vec{i})}, \forall \vec{i}.
\end{align}
Hence, $\ket{\phi_{k}(\vec{i})}$ is an eigenvector of $S_{d,k}$ with eigenvalue $\omega^{q(\vec{i})}$, where $q(\vec{i}) =  -1/m_{k,n} \sum_{j=1}^{n-1} m_{k,j} i_j \ \mod \ d$, i.e. iff 
\begin{align}
\label{eq:EState}
\ket{\phi_{k}(\vec{i})} = c_k(\vec{i}) \ket{q(\vec{i})_k}
\end{align}
for some complex number $c_k(\vec{i})$. 
Reinserting Eq. (\ref{eq:EState}) into Eq. (\ref{eq:deceig}) we see that 
\begin{align}
 \ket{\psi} = \sum_{\vec{i}} c_k(\vec{i}) \ket{\vec{i}_k}\ket{q(\vec{i})_k}.
\end{align}
Hence, $\ket{\psi}$ is a superposition of tensor products of eigenstates of $S_{d,k}$ with the following property. The state of any $n-1$ subsystems 
determines the state of the remaining subsystem.
This implies that the outcomes of measurements of $\{\ket{i}_k\}$ on any $n-1$ subsystems
determine the outcome of a measurement performed in the same basis on the last 
subsystem. At the same time, the outcome of the measurement on one subsystem is completely random if one is unaware of the outcomes of the other measurements,
as the single-subsystem reduced states of $\ket{\psi}$ are completely mixed. 
Considering both of these facts, we obtain 
\begin{align}
\label{eq:MutInfPauli}
 &I(S_{d,k}^{(l)}: S_{d,k}^{(\bar{l})}|\ket{\psi}\bra{\psi}) = \log(d),
\end{align}
for all $l \in \{1,\ldots,n\}$. As this holds for all $k \in \mathcal{K}$ and the eigenbases of the corresponding $N = |\mathcal{K}|$ generalized Pauli operators are mutually unbiased, this shows that
\begin{align}
 \mathcal{C}_N(\psi) = \log(d),
\end{align}
which is the maximal possible value. \qed
}\\

Note that it is straightforward to generalize this theorem in such a way that it also includes other local unitary symmetries $U = U_1 \otimes \ldots \otimes U_n$ for which the spectrum of the local unitaries $U_i$ is $\{\omega_d^j\}_{j=0}^{d-1}$ for all $i \in \{1,\ldots,n\}$.
Recall also that 
any state that maximizes $\mathcal{C}_N$ also maximizes $\mathcal{C}_{N'}$ for $N' \leq N$. 

There are many interesting states that meet the prerequisits of Theorem \ref{thm:2}, as we show in the next section. In the proof of Theorem \ref{thm:2} it became evident that
having a symmetry 
$\bigotimes_{l=1}^n S_{d,k}^{m_{k,l}} \in S_{\psi}$ and having all single-subsystem reduced states completely mixed implies Eq. (\ref{eq:MutInfPauli}), i.e. measurements on $n-1$ subsystems in the corresponding basis determine the 
outcome of the last subsystem. It might be tempting to believe that 
the reverse is true as well for a state $\ket{\psi}$ with completely mixed single-subsystem reduced states. That is, one might think that the existence of such correlations in a basis 
$\{\ket{b_j}\}_{j=0}^{d-1}$ entails the existence of a unitary $U = \sum_{j=0}^{d-1} e^{i \phi_j} \ket{b_j}\bra{b_j}$ such that
\begin{align}
\label{eq:Usymm}
 \bigotimes_{l=1}^n U^{m_l} \ket{\psi} = \ket{\psi},
\end{align}
for some set of integers $\{m_l \neq 0\} \subset \N$.
That this is not the case can be seen by the example of the four-qutrit state $\ket{\psi} \propto \ket{0120} + \ket{1201} + \ket{2012}$. This state is maximally correlated in the 
computational basis in the way described above. It is, however, easy to show that it does not have a nontrivial local symmetry $U$ that fulfills Eq. (\ref{eq:Usymm}) and is diagonal in the computational basis.\\

In what follows, we use the insights gained in this section to study the correlation properties of some multipartite quantum states via the set of correlation functions $\{\mathcal{C}_N\}$ and present applications of these 
measures.

\section{Examples of pure states maximizing $\mathcal{C}_N$}
\label{sec:pure}

In this section we use the theorems and the lemmata proven in the previous section to study the entanglement of multipartite pure states. 
We provide examples of states that maximze $\mathcal{C}_N$ and study the set $\{\mathcal{C}_N\}$ in the context of LOCC transformations.
A state $\ket{\psi}$ meeting the prerequisits of Theorem \ref{thm:2} is interesting in terms of state transformations via LOCC as it has nontrivial local unitary symmetries and completely mixed single-subsystem reduced states. It then follows from the results in \cite{Spee0, deVicente0} that $\ket{\psi}$ is convertible, i.e. it
can be transformed deterministically via LOCC to some other 
(non-LU-equivalent) state in its SLOCC class. In the following we study the entanglement of some of these states and related ones and compare it with $\{\mathcal{C}_N\}$.

\subsection{Three-qubit states}
\label{sec:3qb}

The GHZ state, $\ket{GHZ_{2,3}} \propto \ket{000} + \ket{111}$, is (up to LUs) the unique genuinely tripartite entangled pure 
three-qubit state with completely mixed single-subsystem reduced states. Hence, it is the only pure state that can potentially maximize $\mathcal{C}_N$, according to 
Lemma \ref{lem:1}. Indeed, the maximal value of $\mathcal{C}_2(GHZ) = 1$ can be reached if either $\sigma_x$ or $\sigma_z$ is measured on each of the subsystems.
Note however, that 
$\mathcal{C}_3$ is not maximized if measurements of $\sigma_y$ are included, i.e. if all Pauli operators are measured. In the following we call the scenario where all Pauli operators are measured the Pauli setting. 
As the y-measurement 
does not give rise to any correlations, we have $C_3(GHZ_{2,3},\{\sigma_j^{(l)}\}_{j \in \{x,y,z\},l \in \{1,2,3\}}) = 2/3$. Numerical calculations suggest that in fact $\mathcal{C}_3(GHZ_{2,3}) = 2/3$ and that there is, moreover, no three-qubit pure state that exceeds this value. 
Hence, there appears to be no three-qubit pure state that exhibits correlations in three mutually 
unbiased bases that are strong enough to yield $\mathcal{C}_3 > 2/3$.\\

\subsubsection{States in the maximally entangled set}

$\mathcal{C}_2$ appears to very well capture the multipartite correlations of states in the GHZ class, while its value for the W state, 
$\ket{W} \propto \ket{100}+\ket{010}+\ket{001}$, is comparatively low ($\mathcal{C}_2(W) = 0.685$). 
Due to that reason, we examine states in the intersection of the MES (see Sec. \ref{sec:PrelNot}) with the GHZ class.

Every state in the GHZ class is LU-equivalent to a state of the form \cite{MES}
\begin{align}
\label{eq:GHZMES}
 \ket{\psi_{GHZ}(\vec{g};z)} \propto g_{x_1} \otimes g_{x_2} \otimes g_{x_3}P_z \ket{GHZ_{2,3}},
\end{align}
where $\vec{g} = (x_1,x_2,x_3) \in \R_{\geq 0}^3$, $z \in \C$, $|z| \leq 1$. Moreover, $g_{x_j}$ are invertible operators such that $g_{x_j}^{\dagger}g_{x_j} = 1/2 \one + x_j \sigma_x$, for all 
$j \in \{1,2,3\}$, and $P_z = \text{diag}(z,1/z)$. Note that we choose here $g_{x_j} = \sqrt{1/2 \one + x_j \sigma_x}$. It has been shown in \cite{MES} that the set of states in the GHZ class that are also in the MES is given by the set of states with $z=1$. The GHZ state obviously corresponds to $\ket{\psi_{GHZ}(\vec{0};1)}$. 

\begin{figure}[h!]
 \includegraphics[width=0.5\textwidth]{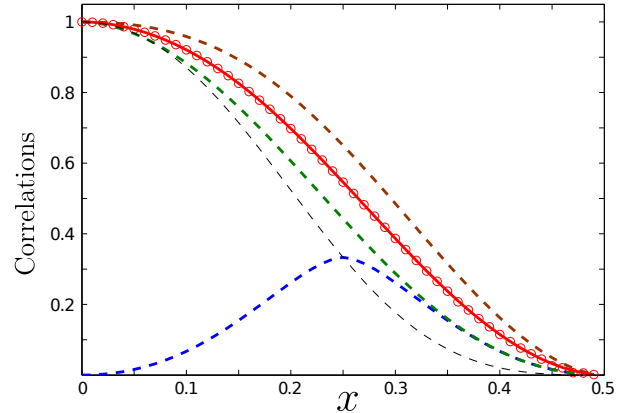}
 \caption{The quantum correlations of the three-qubit state $\ket{\psi_{GHZ}((x,x,x);1)}$, which is an element of the MES, as measured by $\mathcal{C}_2$. 
 The average of the correlations obtained by measuring in the x- 
 and z-basis (solid red line) are equal to the correlations obtained if we optimize
 over all MUBs (red dots) which shows that the optimal choice of MUBs is in fact the former setting. $\mathcal{C}_2$ monotonically decreases with $x$. The correlations in the x-basis, $Q_x$, (uppermost dashed line; red) 
 are larger than the correlations $Q_z$ (middle dashed line; green). $Q_y$  (lowest dashed line; blue) reaches a maximum at $x = 1/4$. The three-tangle, $\tau_3$ (thin grey line; dashed), declines similarily to $\mathcal{C}_2$.}
 \label{fig:1}
\end{figure}

Here, we focus  
on MES states $\ket{\psi_{GHZ}((x,x,x);1)}$ that are symmetric under exchange of subsystems. $\mathcal{C}_2$ decreases monotonically with increasing $x > 0$ as can be seen
in Fig. \ref{fig:1}. This shows that the quantum correlations contained in the GHZ state, as measured by $\mathcal{C}_2$, are particularily strong also in comparison with other states in the MES. 
Let us stress here again that $\mathcal{C}_N$ of a generic state does not need to be optimized for a measurement setting for which
all parties measure the same 
bases. However, for $\ket{\psi_{GHZ}((x,x,x);1)}$ we find numerically that $\mathcal{C}_2(x) = 1/2(Q_x(x) + Q_z(x))$ holds, 
where $Q_i = 1/3 \sum_{l=1}^3 I(\sigma_i^{(l)}:\sigma_i^{(\bar{l})})$ are 
the correlations obtained if $\sigma_i$ is measured on each subsystem, for $i \in \{x,y,z\}$.
The correlations $Q_x(x)$ decrease only slowly,
while $Q_z(x)$ declines more rapidly. 
This is because the local operators that are applied to the GHZ state in order to obtain $\ket{\psi_{GHZ}((x,x,x);1)}$ have, in the Pauli basis, only a component along $\sigma_x$ 
and hence correlations in the x-basis are favoured. Interestingly, the decline of 
$Q_x(x),Q_z(x)$ from their maximal value with increasing $x>0$ is accompanied by the appearance of correlations in the y-basis; $Q_y(x)$ increases from zero 
to a maximal value at $x=1/4$. Clearly all correlations disappear as $x$ approaches the value $1/2$ and $\ket{\psi_{GHZ}((x,x,x);1)}$ 
approaches a separable state. In Fig. \ref{fig:1} we also compare $\mathcal{C}_2$ with the three-tangle, $\tau_3$ \cite{tangle}. Both measures decline in a similar manner with increasing $x$.
However, the three-tangle declines more rapidly.\\

\subsubsection{States in the accessible set of the GHZ state}

\begin{figure}[h!]
 \includegraphics[width=0.45\textwidth]{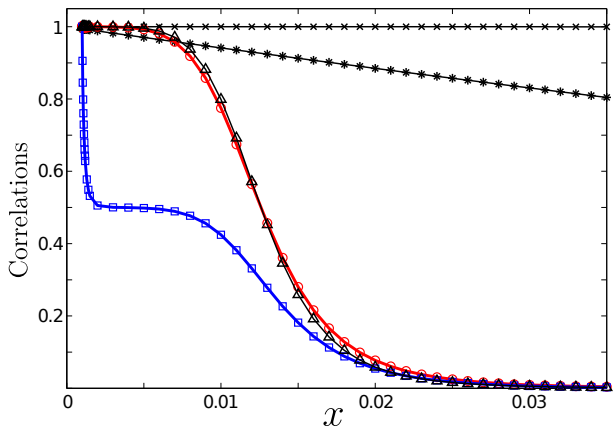}
 \caption{The initial state close to the GHZ state (see main text) is successively transformed via LOCC into states 
 with parameters $\vec{g} = (x,x,x), z(x)$, where $x > 0$ and $z(x)$ is determined by $x$ (see Eq. (\ref{eq:GHZMES})). The uppermost line (black, crosses) depicts the source 
 entanglement and the other straight line (black, asterics) the accessible entanglement of the states. The upper curved line (red, circles) represents 
 $\mathcal{C}_2$ and the lower curved line (blue, squares) the correlations $C_2$ measured in the x- and z-basis (the Pauli setting). 
 Notice that we investigated more intermediate states in the range $x \in [0.001,0.002]$ to 
 better resolve the decline of the correlations in the Pauli setting. The black line with triangles shows the values of $\tau_3$.}
 \label{fig:3}
\end{figure}

The GHZ state is the three-qubit state which is convertible 
to the most other states via LOCC. That is, it has the largest accessible entanglement of all three-qubit states \cite{EnMes}.  
It is interesting to investigate how $\mathcal{C}_2$ changes under LOCC operations. In Fig. \ref{fig:3} we show how these changes compare to the ones of 
the source and the accessible entanglement
(see Sec. \ref{sec:PrelNot}) for a LOCC transformation in which a state with very high $\mathcal{C}_2$ is transformed into
other states along a specific path in Hilbert space. The conditions that the states on such a path have to fulfill have been determined in \cite{turgut, EnMes}. Note that, in order to be very precise, we do not start at the GHZ state itself, as it has 
an accessible entanglement which is not directly comparable to the one of other states (see \cite{EnMes}).
We start at the state $\ket{\psi_{GHZ}((x_0,x_0,x_0);z_0)}$, with $x_0 = 0.001$ and
$z_0 = 0.99999 - 0.00099i$ with a value of $\mathcal{C}_2$ that is very close to 1 (see Fig. \ref{fig:3}). Then the transformation continues to LOCC-reachable states
with larger components $x_j$ for $j \in \{1,2,3\}$, of $\vec{g}$ (see Eq. (\ref{eq:GHZMES})), which also determine their $z$-parameter.
Interestingly, the $z$-parameter deviates a lot from $|z| = 1$ even if $\vec{g}$ is only changed slightly \footnote{See \cite{turgut, EnMes} for details on LOCC transformations in the three-qubit GHZ class.}. As a result, 
the correlations in the x- and z-basis (the Pauli setting) differ significantly from $\mathcal{C}_2$ in the course of the protocol.
However, as the LOCC protocol proceeds, even the optimal setting yields only small correlations. This shows that the quantum correlations measured 
by $\mathcal{C}_2$ can be used up rapidly in the process of a LOCC transformation, while the source and the accessible entanglement decline much slower. Interestingly, the correlations 
measured by the three-tangle, $\tau_3$, and $\mathcal{C}_2$ are almost identical for the states considered in this LOCC transformation.

\subsection{Three-qutrit states}
\label{sec:qutrits}

We have seen above that there exists (up to LUs) only one pure three-qubit state, the 
GHZ state, that can potentially exhibit perfect correlations in MUBs as measured by the functions $\{\mathcal{C}_N\}$. However, we have numerical evidence that even the GHZ state of three qubits can maximize only $\mathcal{C}_2$, but not $\mathcal{C}_3$. 
In contrast to that, we show now that, by increasing the local dimension, one can find infinitely many three-qutrit states that exhibit these 
perfect correlations in four MUBs. We consider the states  
\begin{align}
 \ket{\Psi_{3,3}(a,b,c)} &= a (\ket{000} + \ket{111} + \ket{222}) \label{eq:33state}\\
 &+ b (\ket{012} + \ket{201} + \ket{120}) \nonumber \\
 &+ c (\ket{021} + \ket{210} + \ket{102}), \nonumber
\end{align}
where $a,b,c \in \C$. It is straightforward to show that \cite{hebenstreit}
\begin{align}
 \{S_{3,k}^{\otimes 3}\}_{k \in \{0,1,2\}^2} \subset S_{\Psi_{3,3}(a,b,c)}. \label{eq:33symm}
\end{align}
Note that the seed states of generic three-qutrit SLOCC classes correspond to a subset of these states for which 
$\{S_{3,k}\}_{k \in \{0,1,2\}^2} \equiv S_{\Psi_{3,3}(a,b,c)}$ \cite{hebenstreit}. 
These states are the representatives of the SLOCC classes that are dense in the set 
of three-qutrit pure states \cite{briand}.

\begin{figure}[h!]
 \includegraphics[width=0.5\textwidth]{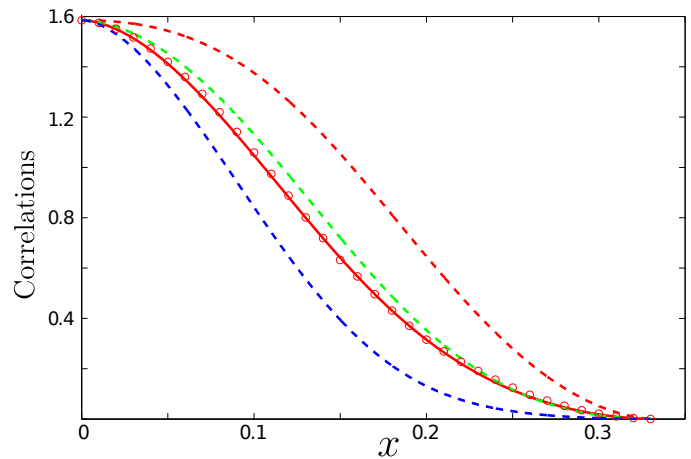}
 \caption{Correlations of the three-qubit state $\ket{\psi((x,x,x);(1,0);1,0,0)}$, which is an element of the MES. The correlations obtained by measuring in the Z-, X-, XZ- and XZZ-basis  
 (solid red line) are in good agreement with the correlations obtained if we optimize
 over the mutually unbiased bases (red dots). $\mathcal{C}_4$ monotonically decreases with $x$. The correlations in the X-basis, $Q_X$, decline slower (uppermost dashed line; red) compared with 
 the correlations $Q_Z$ (middle dashed line; green). The correlations $Q_{XZ}$ and $Q_{XZZ}$ are identical and the weakest (lowest dashed line; blue).}
 \label{fig:2}
\end{figure}

According to Theorem \ref{thm:2}, the fact that $\ket{\Psi_{3,3}(a,b,c)}$ have completely mixed single-subsystem reduced states and the existence of the local symmetries 
in Eq. (\ref{eq:33symm}) imply that $\mathcal{C}_N(\Psi_{3,3}(a,b,c)) = \log(3)$ for $N \in \{2,3,4\}$. In particular, generic three-qutrit 
seed states maximize all $\mathcal{C}_N$. Analogously to the three-qubit case, we can again consider the correlations of states in the intersection of the MES and the SLOCC class represented by 
$\ket{\Psi_{3,3}(a,b,c)}$. In contrast to the three-qubit case, most states in the three-qutrit MES are not convertible via LOCC \cite{hebenstreit}. The convertible states, however, are the only interesting ones 
in terms of LOCC transformations.
For three qutrits they can be expressed as  \cite{hebenstreit}
\begin{align}
  \ket{\psi(\vec{g};k;a,b,c)} = g^{(1)}_k \otimes g^{(2)}_k \otimes g^{(3)}_k\ket{\Psi_{3,3}(a,b,c)},
\end{align}
where $\vec{g} = (g^{(1)},g^{(2)},g^{(3)}) \in \C^3$, $k \in \{0,1,2\}^2$ and $g^{(j)}_k \in \text{span}\{\one, S_{3,k}, S_{3,-k}\}$ such that ${g^{(j)}_k}^{\dagger}g^{(j)}_k = 1/3\one + g^{(j)} S_{3,k} + {g^{(j)}}^* e^{-i \nu_{-k}} S_{3,-k}$. 
Here, the phases $\nu_{k}$ are such that $S_{3,k}^{\dagger} = e^{i \nu_{k}} S_{3,-k}$. Moreover, $g^{(j)}_k \not\propto \one$, except for $\ket{\Psi_{3,3}(a,b,c)}$ itself. We consider states with $\vec{g} = (x,x,x) \in \R_+^3$. Recall that  
$\ket{\Psi_{3,3}(a,b,c)}$ exhibits, in contrast to the three-qubit GHZ state, the same strong correlations for measurements in all of the generalized Pauli bases. We will therefore not see any qualitative
differences in the investigation of $\mathcal{C}_N(\psi(\vec{g};k;a,b,c))$ for different values of 
$k$ and hence choose $k = (1,0)$. In order to simplify the comparison with the qubit case, we moreover choose $a=1, b=c=0$ and thereby states in the SLOCC class of the generalized GHZ state, $\ket{GHZ_{3,3}}$. In Fig. \ref{fig:2} we see, similar to the three-qubit case, that
$\mathcal{C}_4(\psi((x,x,x);(1,0);1,0,0))$ decreases with increasing $x$. We call the bases of 
$S_{(0,1)}, S_{(1,0)}, S_{(1,1)}$ and $S_{(1,2)}$ the Z-, X-, XZ- and XZZ-basis, respectively, and denote the correlations in the basis $W$ by 
$Q_W = 1/3\sum_{l=1}^3 I(W^{(l)}:W^{(\bar{l})})$. The correlation among the measurements in the X-basis remains high even for large values of $g$, while the correlations in the other mutually unbiased bases decrease faster. In contrast 
to the three-qubit case, all of these correlation functions are strictly monotonic functions of $x$.

\subsection{States of more than three subsystems}
\label{sec:moreparties}

There are also genuinely multipartite entangled states $\ket{\psi} \in (\C^{d})^{\otimes n}$, with $n>3$, that reach the maximal value of $\mathcal{C}_N(\psi) = \log(d)$ for some $N$. Note that, for example, all generic four-qubit
seed states have completely mixed single-qubit reduced states and their stabilizer is 
$\{\sigma_i^{\otimes 4}\}_{i \in \{0,x,y,z\}}$, where 
$\sigma_0 \equiv \one$ (see e.g. \cite{MES}). Therefore, they maximize $\mathcal{C}_2$ and $\mathcal{C}_3$ according to Theorem \ref{thm:2}. The same holds for a class of $2^m$-qubit states ($m \geq 2$) with stabilizer 
$\{\sigma_i^{\otimes 2^m}\}_{i \in \{0,x,y,z\}}$ presented in \cite{deVicente0}. Note that also many graph states \cite{hein} maximize $\mathcal{C}_3$. For these states one can construct Pauli symmetries from their stabilizer 
that fulfill the requirements of Theorem \ref{thm:2}. An example is the $n$-qubit GHZ state for $n$ even. Further examples are the states corresponding to the binary tree graphs of eight or of 16 vertices for which 
all leaf nodes have the same depth. Another example is the four-qutrit AME state in Eq. (\ref{eq:AME}), which maximizes $\mathcal{C}_2$ if the generalized Pauli operators $Z_3$ and $X_3$ are measured on each of the subsystems. Going to higher-dimensional systems, it is moreover easy to see that $S_{d,k}^{\otimes n}\ket{GHZ_{d,n}} = \ket{GHZ_{d,n}}$, for all $k \in \{0,\ldots,d-1\}^2$, if $n = m\cdot d$.
Using this, and the fact that the generalized Pauli operators form a complete set of MUBs for prime $d$, it follows from Theorem \ref{thm:2} that $\ket{GHZ_{d,m\cdot d}}$ maximizes $\mathcal{C}_N$ for all $N \in \{2,\ldots,d+1\}$ if $d$ is a prime number. The corresponding measurements are given by the $N$ mutually unbiased generalized Pauli operators.
Another interesting state is the Aharanov state (see e.g. \cite{cabello}), which is defined as 
\begin{align}
 \ket{\mathcal{S}_d} = \frac{1}{\sqrt{d!}} \sum_{i_1,\ldots,i_d = 0}^{d-1} \epsilon_{i_1,\ldots,i_d} \ket{i_1 \ldots i_d} \in (\C^d)^{\otimes d}, \label{eq:Aharanov}
\end{align}
 where $\epsilon_{i_1,\ldots,i_d}$ denotes the generalized Levi-Civit\`{a} symbol. It is easy to see that $\ket{\mathcal{S}_d}$ maximizes the mutual information between 
 any subsystem and the rest, if each of the systems is measured in the computational basis. Moreover, this also holds if any other basis is measured, as the Aharonov state fulfills $U^{\otimes d}\ket{\mathcal{S}_d} = \ket{\mathcal{S}_d}$ for any unitary $U$. This implies that $\ket{\mathcal{S}_d}$ maximizes $\mathcal{C}_N$ whenever $N$ MUBs exist on $\C^d$. In particular, $\ket{\mathcal{S}_d}$ maximizes $\mathcal{C}_2$ for any $d$ and it maximizes $\mathcal{C}_{d+1}$ whenever $d$ is the power of a prime number.\\

The states mentioned so far are all multipartite entangled. There are, however, biseparable states of more than three subsystems that maximize $\mathcal{C}_N$. It is straightforward to see that for a system with Hilbert space 
$\mathcal{H} = (\C^d)^{\otimes n}$, where $n$ is even, any $n/2$ pairs of maximally entangled states maximize $\mathcal{C}_2$. Note, however, that such a biseparable pure state $\ket{\psi_{bisep}}$  can be easily distinguished from genuinely multipartite pure states as they obviously fulfill $S(\rho_A) = 0$ for some genuine subsystem $A \subsetneq  \{1,\ldots,n\}$, where $\rho_A = \text{tr}_{\{1,\ldots,n\}\backslash A}(\ket{\psi_{bisep}}\bra{\psi_{bisep}})$.

\section{Detection of mixed state entanglement}
\label{sec:mixed}

In this subsection we show how $\{\mathcal{C}_N\}$ can be used to detect entanglement of multipartite mixed states of systems with arbitrary dimensions. We focus here on the systems with local dimension $d$. However, the 
results can be easily generalized to systems with different local dimensions.
We first show that any state of a $n$-partite system 
exceeding a certain value of $\mathcal{C}_N$ has to be entangled, which generalizes the results on bipartite entanglement detection 
presented in \cite{CompCorr,huang}. For tripartite systems with Hilbert space $\mathcal{H} = (\C^d)^{\otimes 3}$ we moreover show that this entanglement has 
to be tripartite in nature if $\mathcal{C}_N$ of the corresponding state exceeds another (higher) threshold value. Hence, $\{\mathcal{C}_N\}$ can be used to detect genuine tripartite entanglement using a number of 
local measurements 
in MUBs that scales at most linearly with the local dimension. This is in contrast to the fact that the number of measurements required for entanglement 
detection generally grows rapidly with the size of the system \cite{guehne0}. In \cite{Spengler, erker} related approaches to the detection of multipartite entanglement via MUBs that are not 
based on classical information measures were proposed. We compare our results to those and show that, even if only two local measurement settings are used, 
i.e. if a lower 
bound on $\mathcal{C}_2$ is measured, our method is comparably useful for detecting also high-dimensional tripartite entanglement. Another advantage of the detection 
method presented here is that one does not need to know the 
phase relation between the different measurements (see e.g. \cite{erker}).\\

A state of a system composed of $n$ different $d$-level systems is called fully separable if it is expressible as 
$\rho_{sep} = \sum_j p_j \rho_j^{(1)} \otimes \rho_j^{(2)} \otimes \ldots \otimes \rho_j^{(n)}$, where $\rho_j^{(i)}$ is a state of system $i$, $p_j \geq 0$ and $\sum_j p_j = 1$.
It can be easily seen, as we will explain in the following, that 
\begin{align}
\label{eq:BoundMutInf}
  \sum_{k=1}^N I_{\rho_{sep}}(\mathcal{B}_{k}^{(l)} : \mathcal{B}_{k}^{(\bar{l})}) \leq N \log(d) - f(N,d),
\end{align}
for all parties $l \in \{1,\ldots,n\}$, where $f(N,d)$ is such that the entropic uncertainty relation
\begin{align}
\label{eq:EntrUncert}
\sum_{k=1}^{N} H(\mathcal{B}_k|\rho) \geq f(N,d), \forall \rho \in \mathcal{D}(\C^d),
\end{align}
holds. Note first that the statement in Eq. (\ref{eq:BoundMutInf}) has been proven in \cite{CompCorr} for separable states of bipartite systems in the special case of $N=2$ 
and $f(2,d) = \log(d)$. Note further that any fully separable state is of course separable with respect to any bipartition of the parties. 
It is then easy to see that the proof in \cite{CompCorr} can be generalized to the statement above.\\

Considering the sum over all parties of Eq. (\ref{eq:BoundMutInf}), we obtain that all fully separable states cannot exceed a value of $\mathcal{C}_N = \log(d) - \frac{f(N,d)}{N}$. This 
is equivalent to the following lemma.
\begin{lemma}
\label{lem:fsep}
A state $\rho \in \mathcal{D}((\C^d)^{\otimes n})$ with 
\begin{align}
\label{lem:bisep}
 \mathcal{C}_N(\rho) > \log(d) - \frac{f(N,d)}{N}
\end{align}
is not fully separable, where $f(N,d)$ is such that Eq. (\ref{eq:EntrUncert}) holds.
\end{lemma}

In order for Lemma \ref{lem:fsep} to yield a strong condition for the presence of entanglement, one should obviously choose $f(N,d)$ as large as possible. We have presented examples of $f(N,d)$ in 
Eqs. (\ref{eq:EUR1} - \ref{eq:EUR3}). Lemma \ref{lem:fsep} does not make any statement on 
the nature of the entanglement contained in a state that exceeds the given threshold, e.g. whether it is bipartite, or genuinely multipartite. 
For tripartite systems, however, we provide another condition that is sufficient to show that this 
entanglement is in fact multipartite in nature.

\begin{lemma}
\label{lem:bisep}
A state $\rho \in \mathcal{D}((\C^d)^{\otimes 3})$ with 
\begin{align}
 \mathcal{C}_N(\rho) > \log(d) - \frac{f(N,d)}{3N}
\end{align}
is genuinely tripartite entangled. Here, $f(N,d)$ is again such that Eq. (\ref{eq:EntrUncert}) holds.
\end{lemma}
We provide the proof of Lemma \ref{lem:bisep} in Appendix B.

Using Lemma \ref{lem:fsep} and Eqs. (\ref{eq:EUR1} - \ref{lem:bisep}), i.e. $f(2,d) = \log(d)$, we have that a state $\rho \in  \mathcal{D}((\C^d)^{\otimes n})$ with 
\begin{align}
\label{eq:2measent}
 &\mathcal{C}_2(\rho) > 1/2 \log(d),
\end{align}
contains entanglement, i.e. is not fully separable, and that a state $\rho \in \mathcal{D}((\C^d)^{\otimes 3})$ with 
\begin{align}
\label{eq:2meastripent}
 &\mathcal{C}_2(\rho) > 5/6 \log(d),
\end{align}
contains tripartite entanglement.
The bound in Eq. (\ref{eq:2measent}) is in fact tight as it is saturated by the classically correlated state $\rho_c = 1/d \sum_{i=0}^{d-1} \ket{i}\bra{i}^{\otimes n}$. Note that one might be able to improve the detection 
threshold in Eq. (\ref{eq:2meastripent}). Using 
Eqs. (\ref{eq:EUR3}) we get $f(N,d) = - N \log{\left( \frac{N + d - 1}{d N} \right)}$ and subsequently that any state with 
\begin{align}
\label{eq:boundsd}
 \mathcal{C}_N(\rho) > \log\left( 1 + \frac{d-1}{N} \right)
\end{align}
is entangled and a tripartite state with
\begin{align}
\label{eq:boundsd2}
 \mathcal{C}_N(\rho) > \frac{2}{3} \log(d) + \frac{1}{3} \log\left(1+\frac{d-1}{N}\right)
\end{align}
is genuinely tripartite entangled. 
Using $f(d+1,d) = \frac{d}{2}\log{\frac{d}{2}} + \left(\frac{d}{2}+1\right) \log\left(\frac{d}{2}+1\right)$ of Eq. (\ref{eq:EUR3}) if $d$ is a power of two and $N = d+1$, these 
thresholds can even be improved if a complete set of MUBs exists for the corresponding system. In Table \ref{tab:1} we present numerical values of these bounds for three-qubit and three-qutrit systems.\\

Note that the thresholds 
in Eq. (\ref{eq:boundsd}) decrease with increasing $N$. This, however, does not necessarily imply that $\mathcal{C}_N$ detects more multipartite entangled states than $\mathcal{C}_{N'}$ if $N > N'$. 
There are, e.g., three-qutrit states that are detected by $\mathcal{C}_3$, but not by $\mathcal{C}_2$. However, we show below that there are also
three-qubit states that are detected by $\mathcal{C}_2$, but not by $\mathcal{C}_3$.

\begin{table}
 \begin{tabular}{c|c|c}
  & $d=2$ & $d=3$ \\
 \hline
 $\mathcal{C}_2(\rho_{sep})$ & $1/2$ & $1/2 \log(3)$\\
 $\mathcal{C}_3(\rho_{sep})$ & $1/3$ & $\approx 0.465 \log(3)$\\
 $\mathcal{C}_4(\rho_{sep})$ & $--$ & $\approx 0.366 \log(3)$\\
 $\mathcal{C}_2(\rho_{bisep})$ & $5/6$ & $ 5/6 \log(3)$\\
 $\mathcal{C}_3(\rho_{bisep})$ & $7/9 $ & $\approx 0.822 \log(3)$\\
 $\mathcal{C}_4(\rho_{bisep})$ & $--$ & $\approx 0.790 \log(3)$\\
 \end{tabular}
 \caption{We consider three-qubit ($d=2$) and three-qutrit ($d=3$) systems and depict numerical values of the upper bounds on the value of $\mathcal{C}_N$ that a separable state, $\rho_{sep}$, or a biseparable state, $\rho_{bisep}$, can attain, respectively. If 
 a state yields a higher value, it is detected as entangled or genuinely multipartite entangled, respectively. 
 Recall that the upper bound for $\mathcal{C}_N$ is $1$ for qubits and $\log(3)$ for qutrits. Note that $\mathcal{C}_4$ does not exist for 
 three-qubits as there are at most three mutually unbiased bases on $\C^2$. Note also that in \cite{Riccardi} it has been found numerically that for $d = 3$ and $N = 3$ it holds that $f(3,3) = 3$ (see Eq. (\ref{eq:EntrUncert})) is optimal, 
 and hence the bounds $\mathcal{C}_3(\rho_{sep}) \leq 0.369 \log(3)$ and $\mathcal{C}_3(\rho_{bisep}) \leq 0.790 \log(3)$ hold.}
  \label{tab:1}
\end{table}

\subsection{Examples of detection of low-dimensional tripartite entanglement}
\label{sec:3qbmixed}

In this subsection we give examples of how the functions $\{\mathcal{C}_N\}$ can be used to detect tripartite entanglement in low-dimensional systems.\\

\subsubsection{Detection of genuine three-qubit entanglement}

Let us use $\{\mathcal{C}_N\}$ to detect genuine three-qubit entanglement. 
We have shown in Sec. \ref{sec:3qb} that the three-qubit GHZ state is the only pure state that yields the maximal value of $\mathcal{C}_2(GHZ_{2,3}) = 1$ while we have numerical evidence that there is
no three-qubit pure state that exceeds 
$\mathcal{C}_3(GHZ_{2,3}) = 2/3$, which is below the detection threshold (see Table \ref{tab:1}). This suggests that $\mathcal{C}_2$ is better suited to detect three-qubit entanglement than $\mathcal{C}_3$, as $\mathcal{C}_2$ detects all states with $\mathcal{C}_2 > 5/6$ as genuinely multipartite entangled. 
Although this does not allow to detect the W state, $\ket{W} \propto \ket{100}+\ket{010}+\ket{001}$  
($\mathcal{C}_2(W) = 0.685$) it can be used to detect mixtures of GHZ-type states. The advantage here, as compared to many other detection methods, is that
only two local measurement settings are required.

In the following we use $\mathcal{C}_2$ to detect genuine tripartite entanglement in mixtures of the GHZ state with white noise, i.e. 
\begin{align*}
\rho_{GHZ_{2,3}}(p) = (1-p) \ket{GHZ_{2,3}}\bra{GHZ_{2,3}} + \frac{p}{8} \one.
\end{align*}
Note that $\rho_{GHZ_{2,3}}(p)$ can be detected as tripartite entangled for $p < 4/7$ using at least four local measurement settings \cite{guehne}.
As the maximal value of $\mathcal{C}_2(GHZ_{2,3}) = 1$ is reached if each of the subsystems is measured in the x- or the z-basis we 
can expect that this measurement setting also gives a good lower bound on 
$\mathcal{C}_2(\rho_{GHZ_{2,3}}(p))$. In fact we see numerically that it is optimal. Using this measurement setting, it is then easy to obtain an analytical formula for $\mathcal{C}_2(GHZ_{2,3})$ and 
show that $\mathcal{C}_2$ can detect entanglement up to $5.94\%$ ($p = 0.0594$) of white noise.\\

\subsubsection{Detection of genuine three-qutrit entanglement}

In Section \ref{sec:qutrits} we have seen that there is a continuous set of pure three-qutrit states maximizing each of the correlation functions in $\{\mathcal{C}_N\}_{N=2}^4$. 
Hence, each of these functions can be used to detect genuine tripartite entanglement in the vicinity of these three-qutrit states. Here, we discuss how $\{\mathcal{C}_N\}$ can be used 
to detect tripartite entanglement in two special three-qutrit states in the presence of white noise. These states are the generalized GHZ state, $\ket{GHZ_{3,3}} = \ket{\Psi_{3,3}(1/\sqrt{3},0,0)}$, and the 
Aharonov state, $\ket{\mathcal{S}_3} = \ket{\Psi_{3,3}(0,1/\sqrt{6},-1/\sqrt{6})}$. Before we do so, we review another method to detect multipartite entanglement
via measurements in mutually unbiased bases proposed in  \cite{Spengler}, with which we compare our detection method.\\

In  \cite{Spengler} a set of correlation functions for states $\rho \in \mathcal{D}((\C^d)^{\otimes d})$ has been introduced that can be used to detect genuine multipartite entanglement. 
The authors first defined another correlation function which 
we express in our notation as 
\begin{align*}
 A_{\mathcal{B}}(\rho) = \sum_{i_1,\ldots,i_d = 0}^{d-1}  |\epsilon_{i_1,\ldots,i_d}| \ p(i_1,\ldots,i_d|\{\mathcal{B},\ldots,\mathcal{B}\}; \rho),
\end{align*}
where $\mathcal{B}$ is a basis and $\epsilon_{i_1,\ldots,i_d}$ is the generalized Levi-Civita symbol \footnote{Note that the general definition introduced in \cite{Spengler} uses different bases for different subsystems. 
However, for the example considered here the definition of $A_{\mathcal{B}}(\rho)$ is used.}. Using this function, they then defined the correlation function  
\begin{align}
 J_N(\rho) = \max_{\{\mathcal{B}_k\}_{k=1}^N} \sum_{k = 1}^N A_{\mathcal{B}_k}(\rho),
\end{align}
where $\{\mathcal{B}_k\}_{k=1}^N$ is a set of MUBs, and showed that 
\begin{align}
\label{eq:Corr1bound}
 J_N(\rho_{bisep}) \leq 1 + \frac{N-1}{d}.
\end{align}
Moreover, they showed that $\ket{\mathcal{S}_3}$ reaches the maximum of $J_N(\mathcal{S}_3) = N$ for $N \in \{2,3,4\}$.\\

Let us now compare the two methods for the GHZ and the Aharanov state mixed with white noise, i.e. for 
\begin{align*}
 &\rho_{GHZ_{3,3}}(p) = (1-p)\ket{GHZ_{3,3}}\bra{GHZ_{3,3}} + \frac{p}{27}\one,\\
 &\rho_{\mathcal{S}_3}(p) = (1-p)\ket{\mathcal{S}_3}\bra{\mathcal{S}_3} + \frac{p}{27}\one.
\end{align*}

For both sets of correlation functions we measure both states in the Pauli setting. As expected, $\mathcal{C}_4$ and $J_4$ detect the most states.
We see that $\rho_{\mathcal{S}_3}(p)$ can be detected by $\mathcal{C}_4$ to be genuinely tripartite entangled up to a value of at least $9.18$\% \footnote{This is a lower bound as we did not optimize over all bases.}. 
Not suprisingly, $\{J_N\}$ 
surpasses this value considerably with a detection threshold of $64.29$\%. However, the set $\{\mathcal{C}_N\}$ is better in detecting the states 
$\rho_{GHZ_{3,3}}(p)$. Using the Pauli measurement setting we obtain a lower bound on $\mathcal{C}_N(\rho_{GHZ_{3,3}}(p))$ for which numerical investigations reveal
that it is in fact tight, i.e. that the optimal MUBs are the eigenbases of the generalized Pauli operators, $\{Z_3,X_3,X_3Z_3,X_3Z_3^2\}$. The corresponding analytical expression reads
\begin{widetext}
\begin{align*}
 &\mathcal{C}_N(\rho_{GHZ_{3,3}}(p)) = \frac{1}{9N}((6 N-4) p \log (p)-3 (N-2) (2 p-3) \log (3-2 p)+(9-8 p) \log (9-8 p))
\end{align*}
\end{widetext}
Comparing this with the detection thresholds in Table \ref{tab:1} we see that $\mathcal{C}_4$ detects more states than $\mathcal{C}_2$ and $\mathcal{C}_3$ and yields a lower bound of $8.33$\% up to which the 
state $\rho_{GHZ_{3,3}}(p)$ is detected to be genuinely 
multipartite entangled. In order to find 
$\{J_N(\rho_{GHZ_{3,3}}(p))\}_{N=1}^4$, we perform a numerical optimization over mutually unbiased bases of three-qutrits. 
Note that we optimize here only over unitary rotations of the generalized Paulis 
such that the obtained threshold is just a lower bound. We find that $\rho_{GHZ_{3,3}}(p)$ is not even detected for $p=0$ by any of the lower bounds on $\{J_N\}$ obtained in this way.\\

\subsubsection{Examples of detection of high-dimensional tripartite entanglement}
\label{sec:highdimentdet}

\begin{figure}
 \includegraphics[width=0.5\textwidth]{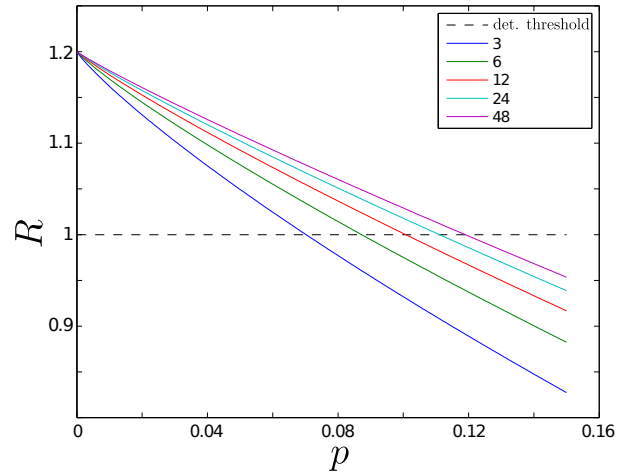}
 \caption{Each of these lines represents $R(p;d)$ (see Eq. (\ref{eq:R})) for some $d$ indicated in the legend. $\rho_{d,3}(p)$ is no longer detected as being genuinely 
 tripartite entangled if the line corresponding to the dimension $d$ is below one. 
 The lower $d$ is, the faster the correlation decreases in relation to the detection threshold. The maximal noise-levels that can be tolerated are  
 $p_{max}(3) = 7.09\%, p_{max}(6) = 8.82 \%, p_{max}(12) = 10.17\%, p_{max}(24) = 11.18\%, p_{max}(48) = 11.95 \%$.}
 \label{fig:High1}
\end{figure}

\begin{figure}
 \includegraphics[width=0.5\textwidth]{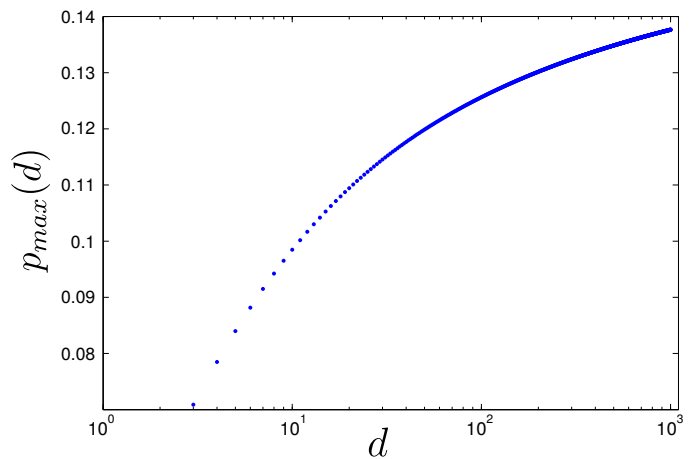}
 \caption{This figure shows $p_{max}(d)$ for $3 \leq d \leq 1000$, i.e. the noise level for which $R(p_{max}(d);d) = 1$ holds (see Eq. \ref{eq:R}). Note the logarithmic scale for $d$.}
 \label{fig:High2}
\end{figure}

We have shown above that $\{\mathcal{C}_N\}$ can be used to detect genuine tripartite entanglement in states of small local dimensions. Here, we show that
Lemma \ref{lem:bisep} also yields nontrivial detection thresholds for states of higher local dimension $d$. Note that 
\begin{align*}
 X_d^{\otimes 3}\ket{GHZ_{d,3}} = Z_d^{d-2}\otimes Z_d \otimes Z_d \ket{GHZ_{d,3}} = \ket{GHZ_{d,3}},
\end{align*}
and hence Theorem \ref{thm:2} implies that $\mathcal{C}_2(GHZ_{d,3}) = \log(d)$ is maximal. The corresponding measurements are performed in the eigenbases of 
$X_d$ and $Z_d$. Hence, $\mathcal{C}_2$ can always be used to detect genuine 
tripartite entanglement in the vicinity of $\ket{GHZ_{d,3}}$. As an example we show here how tripartite entanglement can be detected in 
$\rho_{d,3}(p) = (1-p)\ket{GHZ_{d,3}}\bra{GHZ_{d,3}}+p/d^3 \one$ for nontrivial values of $p>0$. 
Note that it is easy 
to derive an analytical formula for $C_2(\rho_{d,3}(p))$ which is obtained by measuring in the eigenbases of $X_d$ and $Z_d$ and which is a lower bound on $\mathcal{C}_2(\rho_{d,3}(p))$. 
We can certify the presence of tripartite entanglement whenever $C_2(\rho_{d,3}(p))$ is larger than the detection threshold of $5/6\log(d)$ in 
Eq. (\ref{eq:2meastripent}), i.e. whenever the quantity 
\begin{align}
\label{eq:R}
 R(p;d) \equiv \frac{6}{5} \frac{C_2(\rho_{d,3}(p))}{\log(d)} 
\end{align}
fulfills $R(p;d) > 1$. 
Using this formula, one can then 
determine up to which maximal noise level, $p_{max}(d)$, the state $\rho_{d,3}(p)$ can be detected as genuinely tripartite entangled (excluding $p_{max}(d)$ itself), i.e. for which $R(p_{max}(d);d) = 1$.
In Fig. \ref{fig:High1} it is shown how $R(p;d)$ decays as a function of $p$ for some fixed dimensions $d$. The corresponding $p_{max}(d)$ are given in the caption of the figure.
In Fig. \ref{fig:High2} the values of $p_{max}(d)$ are depicted for $3\leq d \leq 1000$. $p_{max}(d)$ grows considerably with increasing $d$.
In fact, it is easy to show that $R(p;d)$ is to leading order independent of $d$ for large $d$, i.e. $R(p;d) \approx 6/5(1-p)$ for $d>>1$.
Due to this scaling, it is straightforward to see that a noise level of around  
$p_{max} = 1/6$ is the most that can still be detected by $C_2$ for large $d$. The values of $p_{max}(d)$ in Fig. \ref{fig:High2} clearly show  
that $\mathcal{C}_2$ can be used to detect even high-dimensional multipartite entanglement 
to relatively high noise levels by using only two local measurements. Note again that it is not necessary to know the phase relations between the locally measured MUBs, which can be 
an advantage in experiment. These results are in contrast to the fact that the complexity associated to the 
detection of entanglement generally grows rapidly with the system size. Note moreover that, for some dimensions $d$ (e.g. $d = 3$), 
the detection efficiency can be increased if more MUBs are included in the measurement, i.e. if $\mathcal{C}_N$ is considered, with $N>2$.\\

\section{Generalization to mutually unbiased measurements}
\label{sec:genmeasurements}

Recently, the authors of \cite{Gour1} generalized the concept of MUBs to mutually unbiased measurements (MUMs). These sets of generalized measurements coincide with MUBs iff their POVM elements are rank one projectors.
It has been shown that the method of entanglement detection via MUBs proposed in \cite{spengler0} can be generalized for bipartite systems to MUMs and 
that this increases the detection strength if no complete set of MUBs is known \cite{chen}. As we show here, the results derived in the previous sections can be generalized to the case where MUMs are measured on the 
subsystems instead of MUBs.
Two generalized measurements in $\C^d$, $\mathcal{P}_1 = \{P_1(i)\}_{i=0}^{d-1}$ and $\mathcal{P}_2 = \{P_2(i)\}_{i=0}^{d-1}$, are called mutually unbiased if \cite{Gour1}
\begin{align*}
 &\text{tr}(P_k(i)) = 1,\\
 &\text{tr}(P_1(i)P_{2}(i')) = \frac{1}{d},\\
 &\text{tr}(P_k(i)P_{k}(i')) = \delta_{i,i'}\kappa + (1-\delta_{i,i'})\frac{1-\kappa}{d-1},
\end{align*}
for all $k \in \{1,2\}$ and for all $i,i' \in \{0,\ldots,d-1\}$, 
where $1/d < \kappa \leq 1$ and $\kappa = 1$ holds iff all POVM elements are rank one projectors and the MUMs are therefore also MUBs. A set of $d+1$ generalized measurements that are mutually unbiased with the same 
$\kappa$ is called a complete set of MUMs. In \cite{Gour1} it has 
also been shown how a complete set of MUMs can be constructed for any $d$ from any orthogonal basis of the space of hermitian, traceless operators that map $\C^d$ to itself.

Due to the operational character of $\mathcal{C}_N$, its generalization to MUMs is obvious. Instead of measuring the classical correlations obtained by measuring in MUBs (see Eq. (\ref{eq:C1}-\ref{eq:C2}) for the 
definition of $\mathcal{C}_N$), we use the same entropic quantities
to quantify the classical correlations obtained by measuring in MUMs. The thresholds 
for entanglement detection in Lemma \ref{lem:fsep} and Lemma \ref{lem:bisep} can also be easily generalized to MUMs. In order to see this, note first that a complete set of MUMs $\{\mathcal{P}_k\}_{k=1}^{d+1}$ fulfills the 
uncertainty relation \cite{Gour1}
\begin{align}
\label{eq:EntMUM}
 \sum_{k=1}^{d+1} H(\mathcal{P}_k|\rho) \geq (d+1)\log\left(\frac{d+1}{1+\kappa}\right).
\end{align}
The inequality in 
Eq. (\ref{eq:EntMUM}) appears to be stronger for smaller $\kappa$. However, the uncertainty of each measurement itself has to be included in any judgement about the mutual unbiasedness of the MUMs. This can be easily seen 
by looking at a measurement with the minimum value of $\kappa = 1/d$, whose measurement operators are completely mixed and therefore do not yield any information about the state of the system. In \cite{Gour1} it has been
shown how one can account for this uncertainty inherent to non-projective measurements in order to get improved uncertainty relations. Using Eq. (\ref{eq:EntMUM}) one can, similar to Lemma \ref{lem:fsep} for MUBs, show that a state $\rho$ that is measured in a set of MUMs with 
efficiency parameter $\kappa$ that yields the correlations $C_{d+1}(\rho)$ is entangled if 
\begin{align}
 C_{d+1}(\rho) > \log(d)-\log\left(\frac{1+d}{1+\kappa}\right)
\end{align}
 and genuinely tripartite entangled if 
\begin{align}
 C_{d+1}(\rho) >  \log(d)-\frac{1}{3}\log\left(\frac{1+d}{1+\kappa}\right).
\end{align}
In future research it would be interesting to investigate whether MUMs can increase the detection efficiency of the presented method for some systems.

\section{Conclusion}
\label{sec:conclusion}
We have presented a set of new measures of multipartite quantum correlations, $\{\mathcal{C}_N\}$.  
$\mathcal{C}_N$ measures multipartite quantum correlations in terms of classical correlations contained in the measurement results of local measurements in $N$ mutually unbiased bases.
This generalizes the approach presented in \cite{CompCorr} to the multipartite realm. We have derived some necessary and some sufficient conditions for a 
multipartite state to maximize $\mathcal{C}_N$. 
These allowed us to show that many interesting multipartite quantum states maximize these measures. Examples are states that have been identified as being particularily important in the context of
LOCC transformations in earlier works; other examples are certain graph states
and generalized GHZ states. We have moreover shown that the set of correlations $\{\mathcal{C}_N\}$ can be used to detect genuine tripartite entanglement using only 
a few local measurement settings in mutually unbiased bases. This allows to detect even high-dimensional tripartite entanglement using only two local measurement settings whose phase relation does not need to be known.

As $\mathcal{C}_N$ is a correlation measure with a clear operational meaning, it would be interesting to see in future research if it can reveal new applications for states that maximize $\mathcal{C}_N$. Moreover, it would be interesting to study the generalization of $\mathcal{C}_N$ to MUMs
in more detail and investigate whether this leads to an increased efficiency in entanglement detection. Finally, it would be appealing to investigate how the correlation measures $\{\mathcal{C}_N\}$ could be used to detect 
genuine multipartite entanglement of systems comprised of more than three subsystems.

\begin{acknowledgments}
The research of DS and BK was funded by the Austrian Science Fund (FWF) through grants Y535-N16 and DK-ALM:W1259-N27.
\end{acknowledgments}

\begin{appendix}

\section{Proof of Lemma 1}
In this Appendix we present the proof of Lemma 1. Note that all definitions that are relevant for this 
lemma are stated before Lemma 1 in the main text. For the sake of brevity we do not restate them here. However, we do restate the lemma (see also \cite{CompCorr,coles}).\\

\noindent {\bf Lemma 1.} \hspace{-0.35cm}\\
\emph{
 For a bipartite state $\rho_{AB} \in \mathcal{D}(\C^{d'}\otimes \C^{d})$, with $d' \geq d$, there exist measurements as described before Lemma 1 in the main text for which 
 $I(X:Y_X) + I(Z:Y_Z) = 2 \log(d)$ iff $\rho_{AB}$ admits the decomposition  
\begin{align}
 \rho_{AB} = \sum_k \tilde{q}_k (V_k \otimes \one)\ket{\phi^+}\bra{\phi^+}(V_k^\dagger \otimes \one), \label{eq:LOtomaxentappendix}
\end{align}
where $\{\tilde{q}_k\}$ are probabilities, $V_k$ are isometries mapping $\C^d$ to $\C^{d'}$ and where $V_k$ and $V_{k'}$ have orthogonal images, i.e. $V_k^{\dagger}V_{k'} = 0$, for $k \neq k'$.
}\\

\proof{
In the main part we have already discussed the ``if''-part. Here, we present a proof of the ``only if''-part. 
As noted in the main text, it follows directly from Holevo's Theorem that $I(X:Y_X) + I(Z:Y_Z) = 2 \log(d_B)$ can only be achieved if
\begin{align}
\label{eq:compmixed}
 S\left(\rho^{(B)}\right) = \log(d)
\end{align}
and
\begin{align}
\label{eq:pure}
 S\left(\rho^{(B)}_{R}(i)\right) = 0, \ \forall i \ \text{and for} \ R \in \{X,Z\},
\end{align}
where $\rho^{(B)} = \text{tr}_A(\rho_{AB}) = \sum_{i} p_{R}(i) \rho^{(B)}_{R}(i)$ for $R \in \{X,Z\}$.

For the sake of readability we continue to use the variable $R \in \{X,Z\}$ in all statements that hold 
for both $X$ and $Z$. Eq. (\ref{eq:pure}) implies that $\rho^{(B)}_{R}(i) = \ket{R(i)}\bra{R(i)}$ for some pure states $\{\ket{R(i)}\}$. Using that the measurements of Bob are complete, it is easy to see that Eq. (\ref{eq:noinfo}) in the 
main text implies that Bob's measurements have exactly $d$ POVM elements, $\{Q_{R_y}(i)\}_{i=0}^{d-1}$. In order to be able to perfectly discriminate the states $\{\ket{R(i)}\}$ (and extract the maximal information) and to fulfill Eq. (\ref{eq:noinfo}) it has 
to hold that $Q_{R_y}(i) = \ket{R(i)}\bra{R(i)}$, $\forall i$, and that $\{\ket{R(i)}\}_{i=0}^{d-1}$ is an orthonormal basis. Moreover, the bases $\{\ket{X(i)}\}$ and $\{\ket{Z(i)}\}$, which we call the X- and Z-basis, respectively, have to be mutually unbiased.

Alice's measurements are composed of the POVM elements $\{P_R(j)\}$. If these are more than $d$ POVM elements, some outcomes have to yield the same state on Bob's system. We denote the set of all outcomes $j$ that 
lead to the state $\ket{R(i)}$ on Bob's system by $\mathcal{I}(i)$ and define the operator $\tilde{P}_R(i) = \sum_{j \in \mathcal{I}(i)} P_R(j)$. 
Note that $\sum_{i=0}^{d-1} \tilde{P}_R(i) = \one_A$, where $\one_A$ denotes the identity on the support of $\rho^{(A)} = \text{tr}_B(\rho_{AB})$. 
Eq. (\ref{eq:pure}) implies that 
\begin{align}
\label{eq:sumop}
 \text{tr}_A((\tilde{P}_{R}(i) \otimes \one) \rho_{AB}) = p_{R}(i) \ket{R(i)}\bra{R(i)},
\end{align}
for  all $i \in \{0,\ldots,d-1\}$. 
From Eq. (\ref{eq:compmixed}) it follows that $\text{tr}_A(\rho_{AB}) = 1/d \one$ and using $\sum_i \tilde{P}_R(i) = \one$ we see that 
\begin{align}
 \text{tr}_A(\rho_{AB}) = \sum_{i=0}^{d-1} p_R(i) \ket{R(i)}\bra{R(i)} = \frac{1}{d}\one, \label{eq:isbasis}
\end{align}
which implies that $p_{R}(i) = 1/d, \forall i$.

Let us now write 
\begin{align}
\label{eq:rhoAB}
 \rho_{AB} = \sum_k \ket{\phi_k}\bra{\phi_k},
\end{align}
 where 
$\{\ket{\phi_k}\}$ are orthonormal, unnormalized pure states.
It is easy to see that Eq. (\ref{eq:sumop}) implies that
\begin{align}
\label{eq:phik}
 \text{tr}_A(\tilde{P}_R(i) \otimes \one \ket{\phi_k}\bra{\phi_k}) = q_{R,k}(i) \ket{R(i)} \bra{R(i)}, \forall k,i,
\end{align}
 where $q_{R,k}(i) \geq 0, \sum_k q_{R,k}(i) = p_{R}(i) = 1/d$. 
Let us write $\ket{\phi_k} = B_k \otimes \one \ket{\phi^+}$ for some operator $B_k$.
Using that $\text{tr}_A(W\otimes \one \ket{\phi^+}\bra{\phi^+}) = 1/d \ W^T$ for any operator $W$, it is easy 
to see that Eq. (\ref{eq:phik}) is equivalent to 
\begin{align}
\label{eq:dagger}
B_k^\dagger\tilde{P}_{R}(i)B_k = d \ q_{R,k}(i) \ket{R(i)^*}\bra{R(i)^*}, \ \forall i,k.
\end{align}
Here, $\ket{R(i)^*}$ denotes for all $i \in \{0,\ldots,d-1\}$ the complex conjugate of $\ket{R(i)}$ in the computational basis.
Summing Eq. (\ref{eq:dagger}) over all $i \in \{0,\ldots,d-1\}$ yields
\begin{align}
\label{eq:diag}
 B_k^\dagger B_k = d \sum_{i=0}^{d-1} q_{R,k}(i)\ket{R(i)^*}\bra{R(i)^*}.
\end{align}
As $R \in \{X,Z\}$, Eq. (\ref{eq:diag}) shows that $B_k^\dagger B_k$ is diagonal in both orthonormal bases $\{\ket{Z(i)^*}\}_{i=0}^{d-1}$ and $\{\ket{X(i)^*}\}_{i=0}^{d-1}$. This implies that 
$\{q_{X,k}(i)\}_{i=0}^{d-1} = \{q_{Z,k}(i)\}_{i=0}^{d-1}$ and we set, without loss of generality, $q_{X,k}(i) = q_{Z,k}(i) = q_k(i)$ by resorting the $Z$-basis.

Next, we show that these conditions imply that 
$B_k^\dagger B_k \propto \one$. Recall that $\{\ket{X(i)^*}\}_{i=0}^{d-1}$ and $\{\ket{Z(i)^*}\}_{i=0}^{d-1}$ are MUBs. That is, the unitary 
$U$ with $\ket{X(i)^*} = U \ket{Z(i)^*}$, for all $i$, has entries $U_{lm} = 1/\sqrt{d} e^{-i \phi_{lm}}$, where $e^{-i \phi_{lm}}$ are phases for all $l,m \in \{0,\ldots,d-1\}$. 
Let us now express $B_k^\dagger B_k$ as a matrix in the basis $\{\ket{Z(i)^*}\}_{i=0}^{d-1}$, i.e. $B_k^\dagger B_k = d \text{diag}(q_{k}(1),\ldots,q_{k}(d)) \equiv D_k$. Then Eq. (\ref{eq:dagger}) is equivalent to 
\begin{align}
 B_k^\dagger B_k = D_k = U D_kU^\dagger.
\end{align}
It is straightforward to show that this can only be satisfied if $q_k(i) = q_k(j) = q_k$ for all $i,j$, i.e. if $B_k^\dagger B_k = q_k d \one$. Hence, $B_k = \sqrt{q_k} \sqrt{d} V_k$ for 
some isometry $V_k$ and $\ket{\phi_k} = \sqrt{q_k} \sqrt{d}(V_k \otimes \one) \ket{\phi^+}$. 

Let us now show that $V_k, V_{k'}$ have orthogonal images, i.e. $V_k^{\dagger}V_{k'} = 0$, for $k \neq k'$. Note first that Eq. (\ref{eq:dagger}) is equivalent to 
\begin{align}
\label{eq:dagger2}
 V_k^\dagger\tilde{P}_{R}(i)V_k = \ket{R(i)^*}\bra{R(i)^*}, \ \forall i,k.
\end{align}
Writing $V_k = \sum_{l=0}^{d-1} \ket{\alpha_{R,k}(l)}\bra{R(l)^*}$, where $\{\ket{\alpha_{R,k}(l)}\}_{l=0}^{d-1}$ are orthonormal states of Alice's system. Eq. (\ref{eq:dagger2}) can only be fulfilled if 
$\langle \alpha_{R,k}(l)| \tilde{P}_{R}(i) | \alpha_{R,k}(l')\rangle = \delta_{l,l'} \delta_{l,i}, \forall k.$
This implies that $\tilde{P}_{R}(i)\ket{\alpha_{R,k}(j)} = \delta_{ij} \ket{\alpha_{R,k}(i)}, \forall k$. We can conclude that 
\begin{align}
\label{eq:deltas}
\langle \alpha_{R,k}(i)| \alpha_{R,k'}(j)\rangle = v_{k,k'}^R(i) \delta_{i,j},
\end{align}
where $v_{k,k'}^R(i) \in \C$. In the following we show that $v_{k,k'}^X(i) = \delta_{k,k'}$ holds \footnote{Note that one can of course show $v_{k,k'}^Z(i) = \delta_{k,k'}$ in the same way.}. That $v_{k,k}^X(i) = 1$ holds is obvious. Let us consider the case $k \neq k'$. Inserting $\ket{X(l)^*} = U \ket{Z(l)^*}$ into $V_{k} = \sum_{l=0}^{d-1} \ket{\alpha_{X,k}(l)}\bra{X(l)^*} = \sum_{l=0}^{d-1} \ket{\alpha_{Z,k}(l)}\bra{Z(l)^*}$ it is easy to see that 
$\ket{\alpha_{Z,k}(j)} = 1/\sqrt{d} \sum_{l=0}^{d-1} e^{i \phi_{lj}} \ket{\alpha_{X,k}(l)}$ (and similar for $k'$). Inserting this into Eq. (\ref{eq:deltas}) for $i = 0 \neq j$ it is then easy to see that 
\begin{align}
\label{eq:zeros}
 \langle \alpha_{Z,k}(0)|\alpha_{Z,k'}(j)\rangle = 1/d \sum_{l=0}^{d-1} e^{-i\phi_{l0}} e^{i\phi_{lj}} v_{k,k'}^X(l) = 0,
\end{align}
for $j \in \{1,\ldots,d-1\}$. From the fact that $\langle \phi_k|\phi_{k'}\rangle = \sqrt{q_k q_{k'}}d \bra{\phi^+}(V_k^\dagger\otimes \one)(V_{k'} \otimes \one) \ket{\phi^+} = \sqrt{q_k q_{k'}} \text{tr}(V_k^\dagger V_{k'}) = 0$ (for $k \neq k'$) and that $q_k, q_{k'} \neq 0$ we obtain that $\sum_{m=0}^{d-1} v_{k,k'}^X(m) = 0$. This equation combined with   
Eq. (\ref{eq:zeros}) is equivalent to 
\begin{align}
\label{eq:syszeros}
 M \vec{v} = \vec{0},
\end{align}
where the entries of the matrix $(M_{jl})_{j \in \{1,\ldots,d\},l \in \{0,\ldots,d-1\}} \in \C^{d\times d}$ are $M_{jl} = 1/\sqrt{d} e^{-i\phi_{l0}} e^{i\phi_{lj}}$ for $j \in \{1,\ldots,d-1\},l \in \{0,\ldots,d-1\}$ and $M_{jl} = 1/\sqrt{d}$ for $j = d, l \in \{0,\ldots,d-1\}$, and 
where the entries of $\vec{v}$ are $v_l = v_{k,k'}^Z(l)$ for $l \in \{0,\ldots,d-1\}$. Using that the phases $1/\sqrt{d} e^{-i\phi_{lj}}$ are the entries of the unitary matrix $U$ it is easy to see that $M$ is unitary as well. Hence, $\vec{v} = \vec{0}$ 
is the only solution of Eq. (\ref{eq:syszeros}). That is, $v_{k,k'}^Z(i) = \delta_{k,k'}$ holds and we obtain from Eq. (\ref{eq:deltas}) that $\langle \alpha_{Z,k}(i)| \alpha_{Z,k'}(j)\rangle = \delta_{k,k'} \delta_{i,j}$. 
This shows that the isometries $V_k$ and $V_{k'}$ have orthogonal images. We can thus conclude that $\rho_{AB}$ can be expressed as 
\begin{align}
 \rho_{AB} = \sum_k \tilde{q}_k (V_k \otimes \one) \ket{\phi^+}\bra{\phi^+} (V_k^\dagger \otimes \one), \label{eq:MixedMaxEnt}
\end{align}
where $\{\tilde{q}_k = q_k d\}$ are probabilities and $V_k$ and $V_k'$ are isometries whose images are orthogonal for $k \neq k'$. This is the decomposition 
described in Eq. (\ref{eq:LOtomaxentappendix}). This completes the proof. \qed
}\\

\section{Proof of Lemma 7}
In this Appendix we present the proof of Lemma 7, which we state here again.\\

\noindent {\bf Lemma 7.} \hspace{-0.35cm}\\
\emph{
A state $\rho \in (\C^d)^{\otimes 3}$ with 
\begin{align*}
 \mathcal{C}_N(\rho) > \log(d) - \frac{f(N,d)}{3N}
\end{align*}
is genuinely tripartite entangled. Here, $f(N,d)$ is again such that Eq. (\ref{eq:EntrUncert}) holds.
}\\

\proof{
We prove the equivalent statement that any biseparable state $\rho_{bisep}$  
has to fulfill 
\begin{align}
\label{eq:boundbisep}
 \mathcal{C}_N(\rho_{bisep}) \leq \log(d) - \frac{f(N,d)}{3N}.
\end{align}
Note that a biseparable state can be expressed as 
\begin{align}
 \rho_{bisep} = \sum_j \sum_{l=1}^3 p_j^{(l)} \rho_j^{(l)} \otimes \rho_j^{(\bar{l})} \in \mathcal{D}((\C^d)^{\otimes 3}),
\end{align}
where  $\rho_j^{(l)}$ is a normalized state of subsystem $l \in \{1,2,3\}$, $\rho_j^{(\bar{l})}$ is a normalized state of the remaining subsystems, $p_j^{(l)} > 0$ and $\sum_{j,l} p_j^{(l)} = 1$. 
Using the definition of $\mathcal{C}_N$ and of the mutual information it is easy to see that 
\begin{align}
 \mathcal{C}_N(\rho_{bisep})\leq \log(d) - \frac{1}{3N}\sum_{k=1}^N \sum_{l=1}^3 H(\mathcal{B}_{k}^{(l)}|\mathcal{B}_{k}^{(\bar{l})};\rho_{bisep}) \label{eq:MutInfProof}
\end{align}
where $\{\mathcal{B}_{k}^{(l)}\}_{k,l}$ with $\mathcal{B}_{k}^{(l)} = \{\ket{b_k(i)}_l\}_{i=0}^{d-1}$ is the set of MUBs that optimize $\mathcal{C}_N(\rho_{bisep})$. 
We now derive an upper bound on the second term on the right-hand-side of Eq. (\ref{eq:MutInfProof}). Notice that 
\begin{align}
\label{eq:CondEntrProof}
 H(\mathcal{B}_{k}^{(1)}|\mathcal{B}_{k}^{(\bar{1})};\rho_{bisep}) = \sum_{i_2,i_3=0}^{d-1} p(i_2,i_3|k) \ H(\mathcal{B}_{k}^{(1)}| \rho_{k}^{(1)}(i_2,i_3)).
\end{align}
Here, $p(i_2,i_3|k) \equiv p(i_2,i_3|\{\mathcal{B}_{k}^{(m)}\}_{m \in \{2,3\}};\rho_{bisep})$ is the probability that parties 2 and 3 obtain outcomes 
$i_2$ and $i_3$ in basis $\mathcal{B}_{k}^{(2)}$ and $\mathcal{B}_{k}^{(3)}$, respectively, and $\rho_{k}^{(1)}(i_2,i_3)$ is the state of party 1 after these measurement outcomes have been broadcasted to this party.
It is easy to see that 
\begin{align*}
 \rho_{k}^{(1)}(i_2,i_3) &= \sum_j  \beta_j(i_2,i_3) \rho_j^{(1)} + \gamma(i_2,i_3) \rho'(i_2,i_3),
\end{align*}
where 
\begin{align}
 &\beta_j(i_2,i_3) = \frac{p_j^{(1)} \bra{b_k(i_2), b_k(i_3)}_{2,3}\rho_j^{(\bar{1})}\ket{b_k(i_2), b_k(i_3)}_{2,3}}{p(i_2,i_3|k)}, \label{eq:beta} \\
 &\gamma(i_2,i_3) = 1 - \sum_j \beta_j(i_2,i_3) \geq 0, \nonumber
\end{align}
and $\rho'(i_2,i_3)$ is a normalized state of the first subsystem. 
Using the concavity of the Shannon entropy we get 
\begin{align}
 &H(\mathcal{B}_{k}^{(1)}| \rho_{k}^{(1)}(i_2,i_3)) \nonumber \\ 
 &\geq \sum_j \beta_j(i_2,i_3) H(\mathcal{B}_{k}^{(1)}|\rho_j^{(1)}) + \gamma(i_2,i_3) H(\mathcal{B}_{k}^{(1)}| \rho'(i_2,i_3)) \nonumber \\
 &\geq \sum_j \beta_j(i_2,i_3) H(\mathcal{B}_{k}^{(1)}|\rho_j^{(1)}). \label{eq:boundEntr} 
\end{align}
The definition of $\beta_j(i_2,i_3)$ in Eq. (\ref{eq:beta}) together with Eq. (\ref{eq:CondEntrProof}) and  Eq. (\ref{eq:boundEntr}) then imply
\begin{align}
 H(\mathcal{B}_{k}^{(1)}|\mathcal{B}_{k}^{(\bar{1})};\rho_{bisep}) \geq \sum_j p_j^{(1)} H(\mathcal{B}_{k}^{(1)}|\rho_j^{(1)}).
\end{align}
Adding up these inequalities for all $k \in \{1,\ldots,N\}$ and using the entropic uncertainty relation Eq. (\ref{eq:EntrUncert}) yields 
\begin{align}
\label{eq:entr1}
 \sum_k H(\mathcal{B}_{k}^{(1)}|\mathcal{B}_{k}^{(\bar{1})};\rho_{bisep}) \geq \alpha_1 f(d,N).
\end{align}
Here, we used the notation $\alpha_1 = \sum_j p_j^{(1)}$. We can get similar expressions for parties 2 and 3. Adding up all the 
corresponding inequalities and using that $\alpha_1 + \alpha_2 + \alpha_3 = 1$
we can therefore conclude that
\begin{align}
 \sum_{k=1}^N \sum_{l=1}^3 H(\mathcal{B}_{k}^{(l)}|\mathcal{B}_{k}^{(\bar{l})};\rho_{bisep}) \geq f(d,N).
\end{align}
Using this inequality in Eq. (\ref{eq:MutInfProof}) we obtain Eq. (\ref{eq:boundbisep}), 
which completes the proof. \qed\\
}
\end{appendix}

\end{document}